\documentclass[aps, prb, twocolumn, showpacs, 10pt, floatfix, superscriptaddress, floatfix, longbibliography]{revtex4-2}

\usepackage{lipsum}  
\usepackage[english]{babel}
\usepackage[utf8]{inputenc}
\usepackage[T1]{fontenc}

\usepackage{amsmath}
\usepackage{braket}
\usepackage{amssymb}
\usepackage[normalem]{ulem}
\usepackage{dsfont}
\usepackage{bm}
\usepackage{bbm}
\usepackage[dvipsnames]{xcolor}
\usepackage{stmaryrd}
\usepackage{bbm}
\usepackage{mathtools}
\usepackage{hyperref}
\usepackage{physics}

\usepackage{graphicx}
\usepackage{dcolumn}
\usepackage{bm}
\usepackage{xcolor}
\usepackage{hyperref}
\hypersetup{
    colorlinks=true,
    citecolor=Green,
    linkcolor=Maroon,
    urlcolor=Blue
    }

\makeatletter
\newcommand{\thickhline}{%
\noalign {\ifnum 0=`}\fi \hrule height 1.2pt
\futurelet \reserved@a \@xhline%
}
\newcolumntype{"}{@{%
\hskip\tabcolsep\vrule width 1.2pt\hskip\tabcolsep}%
}
\makeatother

\begin{document}


\preprint{APS/123-QED}

\title{Driven Non-Unitary Dynamics of Quantum Critical Systems}

\author{Bastien Lapierre}
\affiliation{%
Department of Physics, Princeton University, Princeton, New Jersey 08544, USA%
}%

\author{Pietro Pelliconi}
\affiliation{%
Department of Theoretical Physics, University of Geneva, 24 quai Ernest-Ansermet, 1211 Genève 4, Suisse
}%

\author{Shinsei Ryu}
\affiliation{%
Department of Physics, Princeton University, Princeton, New Jersey 08544, USA%
}%

\author{Julian Sonner}%
\affiliation{%
Department of Theoretical Physics, University of Geneva, 24 quai Ernest-Ansermet, 1211 Genève 4, Suisse
}
\affiliation{%
Jefferson Physical Laboratory, Harvard University, Cambridge, MA 02138, USA
}%

\date{\today}


\begin{abstract}
We investigate the interplay between unitary and non-unitary driven many-body dynamics in (1+1)-dimensional quantum critical systems described by conformal field theory (CFT). By formulating a coherent state approach, we demonstrate that the growth of entanglement entropy and energy can be found analytically for a class of non-unitary driven CFTs, where the evolution alternates between real and imaginary time evolution, the latter corresponding to postselected weak measurements.
We find that non-unitary evolution leads to the emergence of steady states at infinite times for the cases of periodic, quasiperiodic, and random drives. 
In a special class of drives, for mixed initial states, we uncover purification phase transitions that arise as a result of the competition between unitary evolution and weak measurements. We compare the CFT evolution with the corresponding non-unitary dynamics of critical lattice models, finding remarkable agreement. 
\end{abstract}



\maketitle


\section{Introduction}
\label{sec:Introdution}

Isolated quantum many-body systems driven out of thermal equilibrium can host unique nonequilibrium phases of matter. Notable examples include Floquet time crystals~\cite{PhysRevLett.117.090402, PhysRevLett.116.250401}, prethermal phases~\cite{PhysRevX.7.011026} and Floquet topological insulators~\cite{PhysRevX.3.031005}---all of which have been experimentally realized in a variety of settings, from solid-state systems to ultracold atoms~\cite{Wang_2013, PhysRevLett.99.220403, Zhang:2017bwh, Kyprianidis_2021}.
While a typical challenge for driven many-body systems is that they heat up to an infinite temperature state~\cite{PhysRevX.4.041048,PhysRevE.90.012110,PhysRevX.10.011030}, non-unitary time evolution, such as stemming from dissipation, offers a path to realize otherwise unstable dynamical phases~\cite{Diehl_2008,PhysRevB.90.195429, Zhu_2019, Mori_2023}.

Another source of non-unitary evolution comes from monitoring quantum systems. Measurements were recently understood to play a crucial role in designing novel nonequilibrium quantum many-body phases. For instance, it was shown in random circuits with projective measurements that tuning the measurement rate can lead to novel types of phase transitions which are witnessed by quantum information-theoretical quantities, such as entanglement entropy. These measurement-induced phase transitions~\cite{PhysRevB.98.205136, PhysRevB.100.134306, PhysRevX.9.031009, PhysRevB.99.224307, PhysRevB.100.064204, PhysRevX.10.041020, PhysRevB.101.104302} separate a highly entangled volume law phase and a quantum Zeno-like area law phase. 
These phenomena have later been shown to emerge in a wide range of systems, including e.g. non-Hermitian dynamics in the no-click limit~\cite{Biella_2021, Turkeshi_2021,PhysRevLett.126.170503, Le_Gal_2023} and periodic Gaussian circuits under weak measurements~\cite{Granet_2023, PhysRevResearch.6.013131, PhysRevResearch.6.023081}.

Measurements may offer particularly fascinating possibilities for quantum systems entangled over long distances. This happens, for instance, at quantum phase transitions as a consequence of strong quantum fluctuations. Such quantum critical systems possess long-range correlations which are universal, i.e. that are the same irrespective of the underlying microscopic details of the system, and which are described by conformal field theories (CFTs). 
Along this direction, recent works have shown how critical ground states may dramatically be altered by measurements~\cite{PhysRevX.13.021026, PhysRevB.108.165120, PhysRevX.13.041042}, affecting their long-range correlations, and even breaking conformal invariance. 
Beyond equilibrium, CFTs have offered numerous insights into far-from-equilibrium many-body dynamics \cite{Calabrese_2005, Calabrese_2016}, and as part of this broader effort, it is natural to ask how measurements affect the non-equilibrium dynamics of generic quantum critical systems.

\begin{figure}[thbp]
    \centering
        \includegraphics[width=0.48\textwidth]{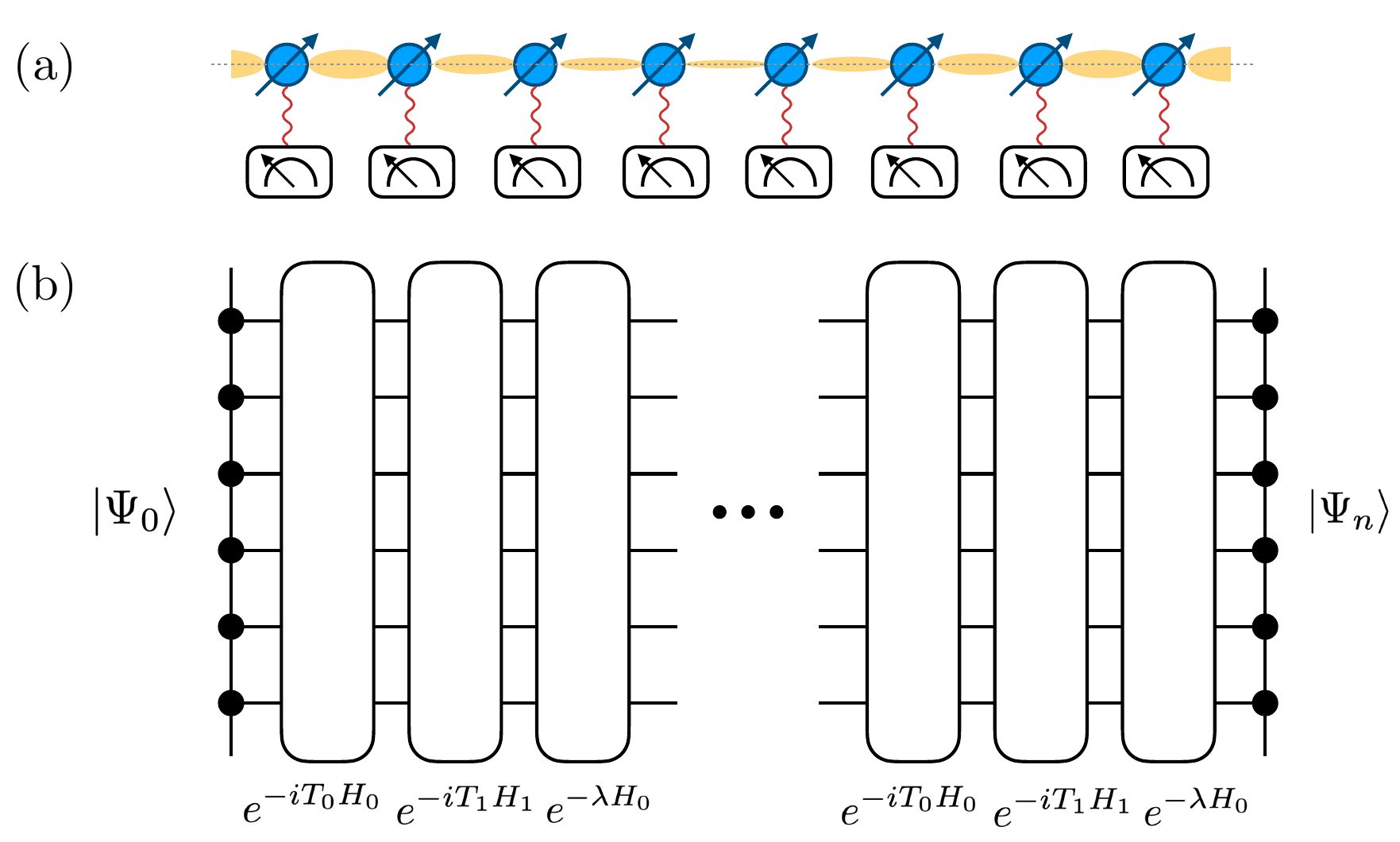}
    \caption{(a) We consider a generic quantum critical system in (1+1)d subject to (i) unitary time evolution with a spatially inhomogeneous Hamiltonian, denoted as $H_1$, for which the couplings of the system vary smoothly over space, and (ii) postselected weak measurements, taking the form of imaginary evolution with the uniform critical Hamiltonian $H_0$. (b) We construct a circuit that alternates between such unitary and non-unitary gates, and prepare a non-equilibrium state after $n$ cycles.  Our goal is to study the dynamics of both pure and mixed states undergoing such evolution for any $(1+1)$d CFT. }
\label{fig:intro}
\end{figure}

In this work, we make a step forward in this direction by studying the dynamics of (1+1)d critical quantum systems evolving under both unitary dynamics and postselected weak measurements. 
Our approach to design such non-unitary evolution is inspired by recent works on the dynamics of driven CFTs~\cite{wen2018floquetconformalfieldtheory, PhysRevX.10.031036, PhysRevResearch.2.023085, 
Han_2020,
PhysRevB.103.224303, PhysRevResearch.3.023044, PhysRevResearch.2.033461, 10.21468/SciPostPhys.13.5.104,10.21468/SciPostPhys.10.2.049, Wen_2022random, de_Boer_2023, lapierre2024floquetengineeredinhomogeneousquantum, Das_2024}, in which multiple dynamical phases were uncovered, such as heating and nonheating phases, corresponding to unbounded linear growth and oscillations in the time evolution of entanglement entropy, respectively.
In order to investigate non-unitary driven dynamics of CFTs, we design postselected weak measurements that crucially do not break the conformal invariance of the underlying critical theory. 
The motivation for this choice is to exploit the universality of CFTs, which in turn allows us to study monitoring in situations where the CFT description is the only one available. In practice, the non-unitary many-body dynamics considered in this paper apply to any (1+1)d CFT, from weakly to strongly coupled.

We demonstrate how the long-time dynamics of driven CFTs is strongly affected by arbitrarily small weak measurements. In most cases, this leads to the emergence of a steady state, independently of the initial state. We show that such phenomenology is robust and arises in Floquet circuits, as well as in aperiodic circuits which alternate either quasiperiodically or randomly between unitary and non-unitary gates.
Strikingly, we analytically show that for a different class of non-unitary Floquet circuits, the interplay between unitary and non-unitary evolution can lead to purification phase transitions~\cite{PhysRevX.10.041020, Li_2021,PhysRevLett.126.060501} for mixed initial states. This transition separates a robust mixed phase and an exponentially purifying phase, both of which can be understood in simple geometric terms as the absence or presence of a stable fixed point in the stroboscopic dynamics. We find the associated critical exponent to be $\frac{1}{2}$, which turns out to be independent of the underlying critical theory. In the non-interacting limit, we numerically show that the non-unitary dynamics of critical lattice models, which could be implemented in quantum simulators~\cite{PRXQuantum.2.010342}, are well described by our field theory results. In the opposite limit of large central charge, we argue that our results open a path to studying the holographic dual of non-unitary many-body dynamics and measurement-induced phase transitions.

The rest of the paper is organized as follows. In Sec.~\ref{sec:Setup}, we first introduce the setup of the paper, as well as our coherent state formalism. In Sec.~\ref{sec:purifications}, this formalism is applied to periodically driven non-unitary CFTs, and then extended to quasiperiodic and random non-unitary evolution in Sec.~\ref{sec:beyondfloquet}. In Sec.~\ref{sec:phasetransitions}, we uncover purification phase transitions for specific classes of periodic non-unitary evolutions. We conclude in Sec.~\ref{sec:Conclusion}, listing some natural extensions and follow-ups.

\section{Setup and approach}
\label{sec:Setup}

\subsection{Setup of the non-unitary circuit}
\label{sec:Setupfirst}

In this work, we consider a class of many-body quantum critical systems in $(1+1)$d described by conformal field theory. In particular, we consider spatially inhomogeneous CFT of finite size $L$~\cite{Allegra_2016, Dubail_2017, Dubail_2017_scipost, Gaw_dzki_2018, PhysRevLett.122.020201, Moosavi_2021} of the form
\begin{equation}
\label{eq:basicmodel}
H = \int_0^L v(x) T_{00}(x) \text{d}x,
\end{equation}
where $T_{00}(x) = \frac{1}{2\pi}(T(x)+\overline{T}(x))$ is the stress-energy tensor of the CFT, and $v(x)$ is a function that varies smoothly over space, encoding the inhomogeneous distribution of energy density. As illustrated in Fig.~\ref{fig:intro}(a), such critical systems arise in the limit of low energies from lattice models with inhomogeneous couplings, leading to lattice Hamiltonians of the form
\begin{equation}
H_{\text{lat}} = \sum_i v_i h_{i},
\label{latticemodeleq}
\end{equation}
where the function $v_i$ varies smoothly over lattice sites, and $h_i$ is the energy density of the model.
The non-equilibrium dynamics following quenches or drives with~\eqref{eq:basicmodel} have recently attracted much theoretical attention and found experimental realizations in quantum simulators~\cite{Tajik_2023}.
It was understood that these non-equilibrium protocols may be exactly solvable in some specific cases~\cite{ PhysRevB.97.184309, Moosavi_2021, PhysRevX.10.031036, PhysRevB.103.224303, PhysRevResearch.3.023044, 10.21468/SciPostPhys.10.2.049}.  This solvability comes from the underlying algebraic structure of \eqref{eq:basicmodel}, since it can be written as a linear combination of the Virasoro generators $\{L_n\}_{n\in\mathbb{Z}}$, and therefore forms an infinite-dimensional algebra. An important simplification occurs when considering a spatial deformation $v(x)$ which only involves a single Fourier mode, $v(x)=\sigma_0+\sigma_+\cos(2\pi q x/L)+\sigma_-\sin(2\pi q x/L)$, where $q\in\mathbb{N}_{>0}$. In this case, the Hamiltonian can be written in terms of only three generators,
\begin{equation}
H = \frac{2\pi}{L}\left[\sigma_0 L_0 + \frac{\sigma_+}{2}(L_q+L_{-q})+\frac{\sigma_-}{2i}(L_{q}-L_{-q})\right],
\label{eq:defham}
\end{equation}
and therefore spans an $\mathfrak{sl}(2,\mathbb{R})$ algebra, a closed subalgebra of the full Virasoro algebra. (Here and in the following, we omit the antichiral component of the stress tensor for simplicity.) As we will show in the next section, the unitary dynamics generated by Hamiltonians of the form~\eqref{eq:defham} is sufficiently simple to retain the exact solvability mentioned earlier, yet giving rise to a variety of dynamical phases, such as heating and nonheating phases.

Our main goal is to study how a source of weak monitoring competes with the above unitary evolution.
On general grounds, dissipation and measurements are expected to break the conformal invariance of the theory, making the analysis of such dynamics extremely challenging. In order to circumvent these limitations, we implement a specific type of postselected weak measurements, which consist of imaginary evolution with the uniform CFT Hamiltonian $H_0\sim\frac{2\pi}{L} L_0$. Indeed, as discussed in App.~\ref{app:postselection}, the non-unitary operator $e^{-\lambda H_0}$ can be understood as a postselected weak measurement of the energy of the critical system, and $\lambda$ is then naturally interpreted as the measurement strength. This choice for the postselection is natural in the context of CFTs, as the stress-energy tensor is a universal operator present for any critical theory. Therefore, we expect this monitoring protocol to naturally apply to any (1+1)d critical theory. 
As will become clear in the remainder of the section, this postselected weak measurement preserves the conformal invariance of the theory, thereby maintaining the solvability of the dynamics.
Importantly, in the case of critical lattice models, this non-unitary dynamics can be efficiently implemented in quantum circuits via coupling to an ancilla~\cite{PRXQuantum.2.010342}.
As a side note, the evolution with a CFT in imaginary time has also been recently investigated in the context of continuously monitored transverse field Ising chains in the no-click limit~\cite{PhysRevResearch.6.013131}.

The resulting non-unitary circuit is depicted in Fig.~\ref{fig:intro}(b): we initialize the system in a state $\ket{\Psi_0}$, typically taken to be the ground state $\ket{0}$ of the Hamiltonian $H_0$, and then alternate between unitary evolutions of the form~\eqref{eq:defham} and weak measurements. 
The initial state thus evolves according to
\begin{equation}
\label{eq:nonunitavprotocl}
\ket{\Psi_0}\mapsto\ket{\Psi_n}= \frac{(e^{-i T_1 H_1}e^{-\lambda H_0})^n\ket{\Psi_0}}{||(e^{-i T_1 H_1}e^{-\lambda H_0})^n\ket{\Psi_0}||}
\end{equation}
in the case of a periodic (Floquet) circuit. We aim to investigate the energy and entanglement entropy of the time-evolved state in the large time limit $n\rightarrow\infty$, and study their universal features, thanks to the underlying conformal invariance of these classes of non-unitary circuits. Concretely, we will study the von Neumann entanglement entropy of the evolved state at large times, 
\begin{equation}
S_A(n) = -\text{Tr}(\rho_A\log\rho_A),
\label{eq:vonneumandef}
\end{equation}
for the reduced density matrix $\rho_A= \text{Tr}_B(\ketbra{\Psi_n}{\Psi_n})$ of a given subsystem $A$.
Alternatively, we will also consider initial mixed states $\rho_0$, and study their purification dynamics at late times in Sec.~\ref{sec:phasetransitions}.

\subsection{Dynamics of coherent states}

We now introduce the main formalism used in this work to carry out the time evolution~\eqref{eq:nonunitavprotocl}. In order to make analytical progress, we will restrict our analysis to a specific class of initial states $\ket{\Psi_0}$ which corresponds to coherent states.
Indeed, the Hilbert space of any $(1+1)$d CFT is constructed by acting with Virasoro generators on the primary states of the theory, which we denote by $\ket{h}$ in the following. Within a given tower of excited states, called Verma module, we can define coherent states as~\cite{perelomov2002coherentstatesarbitrarylie}
\begin{equation}
\ket{\eta, h}=e^{\xi L_{-q}-\bar \xi L_{q}}\ket{h}, \quad \xi = |\xi| e^{i \theta}\in\mathbb{C},
\label{eq:cohernetstate}
\end{equation}
where we have introduced the unit disk coordinate $\eta = \tanh(q|\xi|) e^{i \theta}$ such that $|\eta|<1$ --- each such state is uniquely parametrized by $\eta$. They can also be explicitly written in terms of the orthogonal states $\ket{h,q,n} \propto L_{-q}^n \ket{h}$, and the expression can be found in~\eqref{eq:Coherent_state_explicit_form}. These Virasoro coherent states have attracted considerable attention recently in (1+1)d CFTs, in the context of Krylov and Nielsen complexity~\cite{PhysRevLett.122.231302, caputa2021geometrykrylovcomplexity}, gravity~\cite{Caputa_2023}, and quantum scars~\cite{Liska_2023}.

An important observation is that the unitary evolution with Hamiltonians of the form~\eqref{eq:defham} preserves coherent states of the form~\eqref{eq:cohernetstate}.
 In fact, one can show that (see App.~\ref{app:timeevolucoherent})
\begin{equation}
e^{-i H t}\ket{\eta,h} = \ket{\tilde{\eta},h}, \quad \tilde{\eta} = \frac{a \eta+b}{\bar b\eta+ \bar a}.
\label{eq:transformationlaw}
\end{equation}
where the coefficients $a$ and $b$ may be written explicitly as
\begin{align}
\label{eq:mobiuscoef1}
a &=\cos\left(\frac{q\pi \sqrt{C} t}{L}\right)+i\frac{\sigma_0}{\sqrt{C}}\sin\left(\frac{q\pi \sqrt{C} t}{L}\right), \\
b&=\frac{i(\sigma_+ + i\sigma_-)}{\sqrt{C}}\sin\left(\frac{q\pi \sqrt{C}}{L} t\right),
\label{eq:mobiuscoef2}
\end{align}
with $C= \sigma_0^2-\sigma_+^2-\sigma_-^2$. It is convenient to package the Möbius transformation~\eqref{eq:transformationlaw} as the SU$(1,1)$ matrix
\begin{equation}
    M = \begin{pmatrix}
    a & b \\
    \bar b & \bar a
\end{pmatrix}.
\end{equation}
The advantage of this representation is that successive rounds of evolution applied to the state can be found performing a matrix multiplication between different matrices $M$. The upshot of this discussion is that a transformation on the coherent state is encoded in an SU(1,1) transformation, which leaves the unit disk invariant. Therefore, the unitary evolution maps an initial coherent state, say a primary state $\ket{h}$, to another coherent state. As we shall see later, different types of behavior may emerge in the unitary Floquet dynamics generated by such a family of Hamiltonians: either a fixed point will emerge at the boundary of the disk, leading to an infinite energy state, or no fixed point emerges within the unit disk, leading to oscillations of entanglement entropy and energy.

As discussed in Sec.~\ref{sec:Setupfirst}, our weak measurement protocol consists of imaginary time evolution with the uniform Hamiltonian. The coherent states are also preserved  by such evolution,
\begin{equation}
\label{eq:coherentstatepresr}
\frac{e^{-\lambda H_0}\ket{\eta,h}}{||e^{-\lambda H_0}\ket{\eta,h}||}= |e^{-\frac{2\pi\lambda q}{L}}\eta,h\rangle,
\end{equation}
as shown in App.~\ref{app:timeevolucoherent}.
As expected, in the limit $\lambda\rightarrow\infty$, this projects any coherent state onto their associated highest weight state, i.e. any given $\eta\in\mathbb{D}$ maps to the origin of the disk. In this work, we consider weak measurements of the critical system, thus $\lambda$ is interpreted as a measurement strength which characterizes the coupling of the system to an ancilla. We will then alternate periodically between such weak measurements and unitary evolution, as in Floquet non-unitary circuits~\cite{Granet_2023, PhysRevResearch.6.013131}. Given that both unitary and non-unitary time evolutions map coherent states to coherent states and do not mix between distinct Verma modules, various physical quantities will be directly obtained from equilibrium correlation functions of coherent states~\eqref{eq:cohernetstate}, as we now discuss.

\subsubsection{Energy of coherent states}

A physically relevant quantity to study is the energy of a coherent state, defined as the expectation of the stress tensor $\bra{\eta, h} T(z)\ket{\eta,h}$ (considering for simplicity only the holomorphic sector). In order to compute this correlator, it is instructive to understand the operator $D(\xi)=e^{\xi L_{-q}-\bar \xi L_{q}}$ ($\xi\in\mathbb{C}$) as a displacement operator, in analogy with the coherent states of the harmonic oscillator. In fact, it can be shown that such operators act on primary fields $\mathcal{O}(z)$ of conformal dimension $h$ as
\begin{equation}
D^{\dagger}(\xi)\mathcal{O}(z)D(\xi) = \left(\frac{\partial f}{\partial z}\right)^{h}\mathcal{O}\big( f(\xi,z) \big).
\label{displacementeq}
\end{equation}
To find $f(\xi,z)$, one can write $\xi = |\xi| e^{i \theta}$, and take the derivative of~\eqref{displacementeq} with respect to $|\xi|$, as in~\cite{Liska_2023}. In App.~\ref{app:displacement_primaries} we show that this implies that $f(\xi, z)$ solves the partial differential equation
\begin{equation}
    \Big( e^{-i\theta} z^{1+q} - e^{i\theta} z^{1-q} \Big) \partial_z f - \partial_{|\xi|} f = 0,
    \label{eq:diff_eq_displacement}
\end{equation}
which, demanding that $f(z, 0) = z$, is solved by
\begin{equation}
    f(z)=\left(\frac{z^q-\eta}{1 - \bar{\eta} z^q}\right)^{\frac{1}{q}}.
\label{eq:solution_differential_eq_displacement}
\end{equation}

The energy density, which is a one-point function of a quasi-primary field, can thus be evaluated from the one-point function of the stress tensor in a primary state $\ket{h}$ using~\eqref{displacementeq}, which  leads to~\cite{Caputa_2023, Liska_2023}
\begin{equation}
\bra{\eta, h} T(z)\ket{\eta,h}=\frac{qh_qz^{2(q-1)}(1-|\eta|^2)^2}{(z^q-\eta)^2(1-z^q \bar{\eta})^2}-\frac{c}{24}(q^2-1)\frac{1}{z^2},
\end{equation}
with $z=e^{2\pi i x/L}$ being the spatial coordinate, and $h_q \equiv \frac{1}{q} \Big( h + \frac{c}{24} \, (q^2 - 1) \Big)$, where $c$ is the central charge of the theory.
From this formula, it is clear that the energy density develops $q$ peaks with increasingly high energy in the limit $|\eta|\rightarrow1$; the spatial locations of the peaks are given by solutions to $z^q=\eta$, which reduce to an overall rotation of the $q$-th roots of unity when $|\eta|=1$. This `hotspot' structure in the energy density was previously studied in the context of driven CFTs \cite{PhysRevX.10.031036, PhysRevResearch.2.023085} \footnote{We note that in the context of driven CFTs, the number of peaks in the energy density was found to always be even and equal to $2q$. This comes from the fact that the antiholomorphic part of the CFT was also included, which contributes to another $q$ distinct peaks.}. 
The total energy of a coherent state is then readily obtained by integrating over the spatial coordinate,
\begin{equation}
E(\eta, h)=\frac{2\pi}{L}qh_q\frac{1+|\eta|^2}{1-|\eta|^2}-\frac{\pi c q^2}{12L}.
\label{eq:totaleenetgy}
\end{equation}
In particular, this formula correctly reproduces the energy of a highest weight state $\ket{h}$ for $\eta=0$, and the case $|\eta|\rightarrow1$ corresponds to an infinite energy state, as expected. As we will later explain, Floquet heating will correspond to the flow of time evolution admitting a stable fixed point on the boundary of the disk, leading to a divergent energy evolution.

\subsubsection{Entanglement entropy of coherent states}

The Rényi entanglement entropy of a coherent state $\ket{\eta,h}$ is defined as
\begin{equation}
S_A^{(m)} = \frac{1}{1-m}\log(\text{Tr}(\rho_A^m)),
\end{equation}
for a reduced density matrix $\rho_A= \text{Tr}_B(\ket{\eta,h}\bra{\eta,h})$.  An important particular case is the von Neumann entanglement entropy~\eqref{eq:vonneumandef}, obtained as $\lim_{m\rightarrow1} S_A^{(m)}$.
As is commonplace, the Rényi entropies may be written in terms of correlation functions of twist fields~\cite{Calabrese_2009}
\begin{equation}
S_A^{(m)}(\eta, h) = \frac{1}{1-m} \log \bra{\eta,h}\sigma(z_1)\sigma(z_2)\ket{\eta,h}
\label{eq:Renyi_correlator}
\end{equation}
where $z_{1}$ and $z_2$ correspond to the endpoints of the interval $A=[x_1, x_2]$, and twist fields are primary operators with scaling dimension $h_{\sigma} = \frac{c}{12}(m - \frac{1}{m})$. This correlator is generically theory dependent, and it is not possible to give a universal form valid for any CFT. There is however a special case where this is achievable, which is when the ($t$-channel) operator product expansion (OPE) of~\eqref{eq:Renyi_correlator} is dominated by the identity block. Below, we give a recipe to find the result in this limit. For generic correlators $\bra{\Psi}\sigma(z_1)\sigma(z_2)\ket{\Psi}$ the identity block exchange is given by
\begin{equation}
    \bra{\Psi}\sigma(z_1)\sigma(z_2)\ket{\Psi} = \left[\frac{g'(z_1)g' (z_2)}{(g(z_1)-g(z_2))^{2}}\right]^{h_{\sigma}},
    \label{eq:correlator_with f}
\end{equation}
where $g(z)$ is a solution to the uniformization problem
\begin{equation}
\frac{c}{12}\{g(z),z\} = \bra{\Psi}T(z)\ket{\Psi} ,
\end{equation}
where $\{g(z),z\}=\frac{g'''(z)}{g'(z)}-\frac{3}{2}\left(\frac{g''(z)}{g'(z)}\right)^2$ is the Schwarzian derivative.
While it is in general challenging to solve this uniformization problem, the solution may be found explicitly in the case of $\mathfrak{sl}(2,\mathbb{R})$ coherent states, when $\ket{\Psi} = \ket{\eta, h}$, and is given by~\cite{Caputa_2023}
\begin{equation}
     g(z)=\left(\frac{z^q-\eta}{z^q-\bar{\eta}^{-1}}\right)^{\frac{1}{q}\sqrt{1-\frac{24h}{c}}},
\label{neweq:uniformisation}
\end{equation}
and, in particular, $f$ in~\eqref{eq:diff_eq_displacement} and $g$ in~\eqref{neweq:uniformisation} are related by an overall exponent.
This result for the entanglement entropy is applicable in two distinct limits. First, for generic $h$, only for holographic CFTs, which have a large central charge $c$ and a sparse spectrum of operators $\tilde h$ below the black hole threshold. These contribute to the OPE as $\tilde h/c$, thus suppressed with respect to the identity. We refer to~\cite{Caputa_2023} for a detailed study of the entanglement entropy of coherent states in the holographic limit.
The second limit is when the state is the vacuum, $h=0$, which is the case we will study in the rest of the paper. In this case, the result obtained combining~\eqref{eq:Renyi_correlator},~\eqref{eq:correlator_with f} and~\eqref{neweq:uniformisation} is exact. The reason is that the only block with a non-zero vacuum expectation value in the OPE is the identity. Focusing on this second case, i.e. building coherent states from the CFT vacuum $\ket{0}$, for a half system size interval $A = [0, \frac{L}{2}]$ and for $q$ even, equations~\eqref{eq:Renyi_correlator} and~\eqref{eq:correlator_with f} can be massaged into a simple formula for the von Neumann entanglement entropy $S_A(\eta, 0)$, which is
\begin{equation}
   S_A(\eta, 0) = \frac{c}{3} \log \left(  \frac{| 1 - \eta | \, | (-1)^{q} -  \eta | }{ 1 - | \eta |^2  } \right), \ 
    \label{Half_interval_cft_prediction}
\end{equation}
up to a constant term that we neglect. The usual ground state entanglement entropy is recovered in the limit $\eta \to 0$, where~\eqref{Half_interval_cft_prediction} vanishes. This is expected, since for any $q$ a generic interval $A = [0, \ell]$ has a vacuum entanglement entropy of 
\begin{equation}
   S_A = \frac{c}{3} \log \left[  \sin \left( \frac{\pi \ell}{L} \right) \right] \ , 
\end{equation}
which vanishes when $\ell = L/2$.
Finally, we note that the entanglement scaling of coherent states $\ket{\eta,0}$ still satisfies logarithmic violation of the area law, similarly to the CFT ground state $\ket{0}$, with extra spatial `kinks' that arise due to the lack of translation invariance of such states. These low entanglement features of coherent states are true even for coherent states at arbitrarily high energies, which is consistent with their recent interpretation as quantum scars~\cite{Liska_2023}.

\subsubsection{Overlap of coherent states}

Yet another quantity that is known to characterize non-equilibrium processes is the fidelity, or return probability, defined as $|\langle \Psi(0)|\Psi(t) \rangle|^2$, for an initial state $\ket{\Psi(0)}$, which we will assume to be of the form~\eqref{eq:cohernetstate}. Given that the non-equilibrium time evolution will map any coherent state to another coherent state, it is enough to simply evaluate the overlap between two arbitrary coherent states, $\langle \eta, h|\eta',h' \rangle$. The formula can be obtained explicitly by using the series expansion of coherent states (see App.~\ref{app:timeevolucoherent}), and leads to
\begin{equation}
\langle \eta, h|\eta',h' \rangle = \delta_{h,h'} \left(\frac{1-\eta\bar{\eta}'}{\sqrt{(1-|\eta|^2)(1-|\eta'|^2)}}\right)^{-2h}.
\label{eq:loschmidt}
\end{equation}

\section{Non-unitary Floquet circuits}
\label{sec:purifications}

\subsection{Preliminary: Floquet drives} \label{sec:Unitary_Floquet_drives}

We will study the growth of energy and entanglement entropy for periodically driven CFTs, using the Schrödinger picture when starting from an initial coherent state. This approach differs from previous works on the subject, which used the Heisenberg picture to compute the time evolution of physical observables~\cite{wen2018floquetconformalfieldtheory, PhysRevX.10.031036, PhysRevResearch.2.023085, PhysRevResearch.3.023044}. The advantage of using the Schrödinger picture is that it allows for a natural generalization to non-unitary drives, as considered in the rest of the paper, and will also be necessary to study the evolution of initial mixed states in Sec.~\ref{sec:phasetransitions}.

We consider a Floquet one-cycle unitary operator made of $N$ static steps, taking the general form
\begin{equation}
U_{\text{F}} = U_N
\cdots U_1, \quad U_j = e^{-i H_j T_j},
\end{equation}
where each static segment $H_j$ of the one-cycle evolution is drawn from~\eqref{eq:defham}. This could be for instance seen as the Trotterization of the time evolution with a continuous Hamiltonian $H(t)$.
When the initial state is taken to be a coherent state of the form~\eqref{eq:cohernetstate}, $\ket{\Psi_0}=\ket{\eta, h}$, then the time evolution can be explicitly evaluated as
\begin{equation}
U_{\text{F}}|\eta,h\rangle  = |\tilde{\eta}_1,h\rangle,
\end{equation}
where the new unit disk coordinate is given by the SU(1,1) transformation $\tilde{\eta}_1 = \frac{\alpha_1\eta+\beta_1}{\bar{\beta}_1\eta+\bar{\alpha}_1}$, with
\begin{equation}
\label{onecyclemapseq:}
\begin{pmatrix}
    \alpha_1&\beta_1\\\bar{\beta}_1&\bar{\alpha}_1 
\end{pmatrix} = \begin{pmatrix}
    a_N&b_N\\\bar b_N& \bar a_N 
\end{pmatrix} 
\cdots \begin{pmatrix}
    a_1&b_1\\\bar b_1& \bar a_1 
\end{pmatrix} \in \text{SU}(1,1),
\end{equation}
where each coefficient $a_j$ and $b_j$ are given by~\eqref{eq:mobiuscoef1} and~\eqref{eq:mobiuscoef2}, respectively.
It is therefore clear that the stroboscopic time evolution after $n$ cycles amounts to mapping a coherent state $\ket{\eta,h}$ to another coherent state 
\begin{equation}
\ket{\tilde{\eta}_n,h} = U_{\text{F}}^n \ket{\eta,h},
\end{equation}
such that
\begin{equation}
\tilde{\eta}_n = \frac{\alpha_n\eta+\beta_n}{\bar{\beta}_n\eta+\bar{\alpha}_n}, \quad \begin{pmatrix}
    \alpha_n&\beta_n\\ \bar{\beta}_n & \bar{\alpha}_n
\end{pmatrix} =\begin{pmatrix}
    \alpha_1&\beta_1\\ \bar{\beta}_1 & \bar{\alpha}_1
\end{pmatrix}^n.
\end{equation}
Thus, the stroboscopic time evolution corresponds to a discrete map in the space of coherent states, i.e. in the unit disk $\mathbb{D}$ (we recall that SU(1,1) transformations act as automorphisms of $\mathbb{D}$).
We note that in the language of~\cite{Liska_2023}, where coherent states were identified as quantum scar states of CFTs, our Floquet drive thus acts as a map from one scar to another in the same Verma module.

\begin{figure}[htbp]
    \centering
        \includegraphics[width=0.45\textwidth]{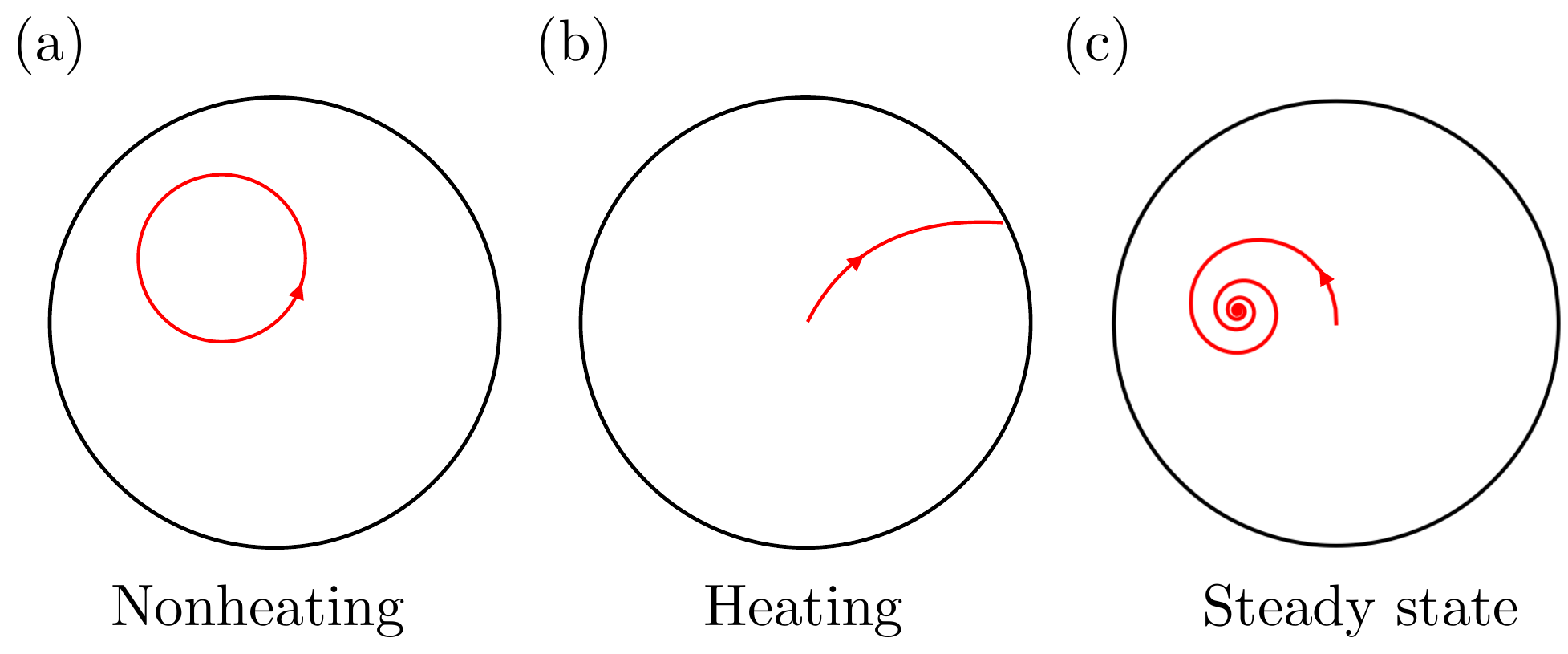}
    \caption{Summary of the Floquet dynamics both in the unitary and the non-unitary case studied in Secs.~\ref{sec:Unitary_Floquet_drives} and~\ref{sec:periodicnonunitary21}. In the unitary setting, two dynamical phases exist can emerge: (a) a nonheating phase, where $\tilde{\eta}_n$ oscillates as function of the stroboscopic time $n$, and (b) a heating phase, where the initial coherent state flows a fixed point $\tilde{\eta}_*$ at the boundary $\partial\mathbb{D}$, corresponding to an infinite energy state. (c) In the non-unitary setting, a steady state necessarily emerges, and any initial coherent state flows to such state, explicitly given by the stable fixed point $\tilde{\eta}_*$ in $\mathbb{D}$.  }
\label{fig:sketch_evolutio}
\end{figure}

It was found in~\cite{wen2018floquetconformalfieldtheory, PhysRevX.10.031036} that depending on the driving parameters, a heating phase and a nonheating phase may arise. In particular, the nonheating phase corresponds to the situation where the one-cycle transformation is elliptic, i.e., $\text{Tr}(M_1)^2<4$, while the heating phase corresponds to a hyperbolic transformation, $\text{Tr}(M_1)^2>4$.
Whether the system is in a heating or nonheating phase thus corresponds to whether the mapping $\tilde{\eta}_n$ has a stable fixed point located at the boundary $\partial \mathbb{D}$ or not, as is illustrated on Fig.~\ref{fig:sketch_evolutio}(a-b). In fact, we can define the stable fixed point associated to the one-cycle SU(1,1) transformation as
\begin{equation}
\label{eq:fixedpointetaff}
\tilde{\eta}_{*} = \frac{\alpha_1 - \bar{\alpha}_1 - \sqrt{\big(\alpha_1 - \bar{\alpha}_1 \big)^2 + 4 |\beta_1|^2}}{2 \bar{\beta}_1}.
\end{equation}
In the heating phase, the stable fixed point $\tilde{\eta}_{*}$ is located on the boundary $\partial\mathbb{D}$, $|\tilde{\eta}_{*}|=1$. Therefore, any coherent state evolves exponentially fast towards such an asymptotic state
\begin{equation}
\lim_{n\rightarrow\infty}\ket{\Psi(n)} = \lim_{n\rightarrow\infty}\ket{\tilde{\eta}_n,h} = \ket{\tilde{\eta}_*,h}.
\end{equation}
This asymptotic state, being located on $\partial \mathbb{D}$, has infinite energy; therefore, any initial coherent state heats up under the Floquet drive in the heating phase. In the nonheating phase, the stable fixed point is located outside of the unit disk, which implies that the dynamics of the map $\tilde{\eta}_n$ stays bounded \textit{within} the unit disk $\mathbb{D}$, see Fig.~\ref{fig:sketch_evolutio}(a). 
Both phases may be diagnosed from the growth of entanglement entropy or energy. In our formalism, the time evolution of energy and (half-system) entanglement entropy after $n$-cycles are given by \footnote{Note that for $S_A(n)$ it is assumed that $h=0$, as explained in Sec.~\ref{sec:Setup}. On the other hand, the formula for the energy holds for any primary state. }
\begin{align}
S_A(n) &= \frac{c}{3} \log \left(  \frac{| 1 - \tilde{\eta}_n | \, | (-1)^{q} -  \tilde{\eta}_n | }{ 1 - | \tilde{\eta}_n |^2  } \right),\\
E(n) &= \frac{2\pi}{L}qh_q\frac{1+|\tilde{\eta}_n|^2}{1-|\tilde{\eta}_n|^2}-\frac{\pi c q^2}{12L}.
\end{align}
We note that these formulas generalize previously found results~\cite{wen2018floquetconformalfieldtheory, PhysRevX.10.031036, PhysRevResearch.2.023085}, where the initial state $\ket{\Psi(0)}$ was assumed to be the CFT vacuum $\ket{0}$. From these formulas, it is clear that in the heating phase the energy (entanglement entropy) grows exponentially (linearly) given that $\tilde{\eta}_n$ converges exponentially to the stable fixed point. Similarly, the fidelity $|\langle \Psi(0)|\Psi(n)\rangle|^2$ can be evaluated directly from the overlap formula~\eqref{eq:loschmidt},
\begin{equation}
|\langle\Psi(0)|\Psi(n) \rangle|^2 =  \left(\frac{|1-\eta_0\bar{\tilde{\eta}}_n|}{\sqrt{(1-|\eta_0|^2)(1-|\tilde{\eta}_n|^2)}}\right)^{-4h}.
\end{equation}
It is clear that for any initial state $\ket{\eta,h}$, the fidelity goes exponentially fast to zero as the coordinate $\tilde{\eta}_n$ converges to $\partial \mathbb{D}$.

As we will explore in the rest of the section, yet another kind of dynamics could in principle emerge: the stable fixed point $\tilde{\eta}_*$ could be located in the interior of the unit disk $\mathbb{D}$. As we will argue, this only happens if the one-cycle map~\eqref{onecyclemapseq:} goes beyond SU(1,1), and is a general element of SL$(2,\mathbb{C})$.

\subsection{Periodic non-unitary circuits}
\label{sec:periodicnonunitary21}

As a next step, we investigate the competition between the unitary Floquet drive described in Sec.~\ref{sec:Unitary_Floquet_drives} and non-unitary evolution in the form of imaginary evolution with the CFT Hamiltonian. In other words, the Floquet operator generically reads
\begin{equation}
U_{\text{F}} = e^{-\lambda H_0}e^{-i T_N H_N}\cdots e^{-i T_1 H_1},
\label{eq:non_unitary_Floquet_drive_protocol}
\end{equation}
where each of the unitary steps is of the form~\eqref{eq:defham}. Intuitively, the unitary part of the drive leads to a linear increase of entanglement when the parameters are taken to be in the heating phase, while the non-unitary part of the evolution tends to bring any coherent states of the form~\eqref{eq:cohernetstate} to the lowest energy state of the Verma module, which significantly reduces entanglement. As a consequence, we expect the interplay between these two processes to stabilize a nontrivial steady state at large times. 

In order to make analytical progress, we note that both the unitary and non-unitary segments of the drive preserve coherent states (see Eqs.~\eqref{eq:transformationlaw} and~\eqref{eq:coherentstatepresr}). Therefore, we have
\begin{equation}
\frac{U_{\text{F}}^n\ket{\Psi_0}}{||U_{\text{F}}^n\ket{\Psi_0}||} = \ket{\tilde{\eta}_n,h},
\end{equation}
where now the transformation that encodes the time evolution of the coherent states generically takes the form $M_{\text{F}}^n$, where
\begin{equation}
M_{\text{F}}=\begin{pmatrix}
    e^{-\frac{\pi\lambda q}{L}}&0\\0& e^{\frac{\pi\lambda q}{L}}
\end{pmatrix}\begin{pmatrix}
    a_N&b_N\\\bar b_N &\bar a_N
\end{pmatrix} \dots \begin{pmatrix}
    a_1&b_1\\\bar b_1 &\bar a_1
\end{pmatrix},
\label{eq:NonUnitary_M}
\end{equation}
which we will later denote as $M_{\rm F} = M_0 M_N \cdots M_1$.
Each matrix in the product belongs to SU(1,1), which maps the unit disk to itself, as discussed previously. However, the non-unitary transformation is an overall scaling and does not belong to SU(1,1), which implies that the product lies within SL(2,$\mathbb{C}$) for nonzero measurement strength. As a consequence, it is clear that the one-cycle transformation does not preserve the unit disk and maps it into a subset of it --- this is intuitively clear since the non-unitary evolution shrinks the unit disk, mapping it into the origin in the limit $\lambda\rightarrow\infty$. The classification of the associated Möbius transformation goes beyond the elliptic, hyperbolic, and parabolic cases studied in the unitary setting, as the trace of $M_{\text{F}}$ is in general complex valued, $\text{Im}(\text{Tr}(M_\text{F}))\neq0$. 
This leads to a loxodromic Möbius transformation, which implies that now a stable fixed point $\eta_*$ appears within the unit disk, instead of its boundary, as seen on Fig.~\ref{fig:sketch_evolutio}(c). The physical consequence is that the steady state is a coherent state with finite energy and entanglement. This is in stark contrast with the unitary setting, where the time-evolved state either oscillates between a set of coherent states or converges towards an infinite energy state. 

\begin{figure}[htbp]
    \centering
        \includegraphics[width=0.48\textwidth]{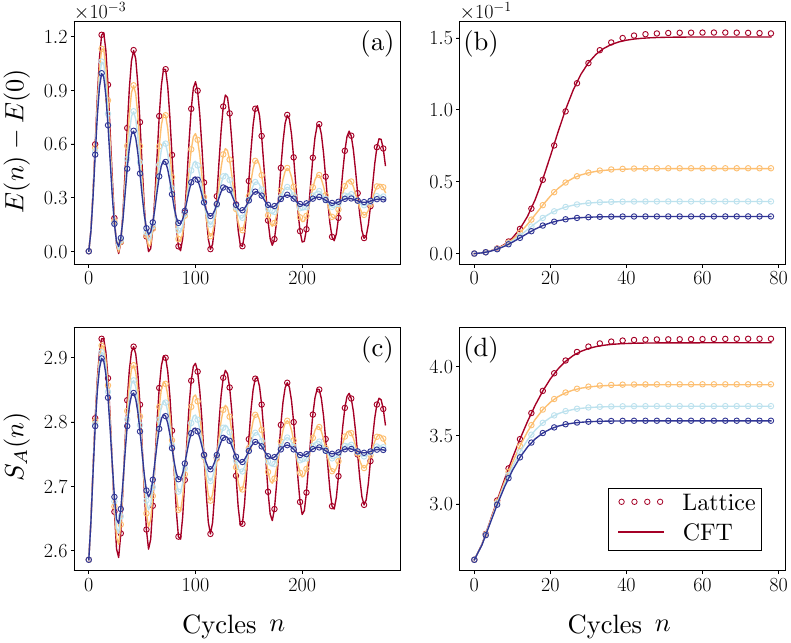}
    \caption{Time evolution of (a-b) the energy $E(n)$ and (c-d) the entanglement entropy $S_A(n)$ for a periodic non-unitary drive of the form $U_{\text{F}} = e^{-\lambda H_0}e^{-iT_0 H_0}e^{-iT_1 H_1}$, both from CFT and lattice calculations. We choose parameters $(\sigma_0, \sigma_+, \sigma_-) = (1, 0, 0)$ in~\eqref{eq:defham} for $H_0$, and $(0, 1, 0)$ for $H_1$. For each plot, the values of $\lambda$ are $\{0.16, 0.4, 0.64, 0.88\}$ (red to blue). We observe two distinct regimes, corresponding to damped oscillations ($T_0/L=1/50$, $T_1/L=1/100$) or growth and then saturation to the steady state ($T_0/L=1/100$, $T_1/L=1/50$). The exact CFT results (solid curves) are compared to free-fermionic lattice calculations (hollow dots) with model~\eqref{eq:latticemodel}, for a system of size $L=800$ and periodic boundary conditions, starting from the ground state of the uniform chain. We observe a near perfect agreement between lattice and CFT data, which improves as the measurement strength $\lambda$ is increased.}
\label{fig:lattice}
\end{figure}

As we argue in App.~\ref{app:fixepoints}, there always exists a stable fixed point within the unit disk as long as $\lambda\neq0$. This implies that the phase transition of the unitary Floquet drives gets washed out by any infinitesimally small measurement rate $\lambda$.
Nevertheless, there are two distinct behaviors in the time evolution of energy and entanglement entropy at intermediate times: in one case, both quantities display damped oscillations around the steady state, while in the other case, there is a linear (exponential) growth of entanglement entropy (energy) before reaching the steady value. These are illustrated on Fig.~\ref{fig:lattice} for a non-unitary two-step drive of the form $U_{\text{F}} = e^{-\lambda H_0}e^{-iT_0 H_0}e^{-iT_1 H_1}$. The value of the steady state half-system entanglement entropy and energy are given by
\begin{align}
S_{A, \infty} &= \frac{c}{3} \log \left(  \frac{| 1 - \tilde \eta_{*} | \, | (-1)^{q} -  \tilde \eta_{*} | }{ 1 - | \tilde \eta_{*} |^2  } \right),\\
E_{\infty} &= \frac{2\pi}{L}qh_q\frac{1+|\tilde \eta_{*}|^2}{1-|\tilde \eta_{*}|^2}-\frac{\pi c q^2}{12L}.
\end{align}
The non-unitary circuit~\eqref{eq:non_unitary_Floquet_drive_protocol} provides a powerful protocol to engineer any coherent state~\eqref{eq:cohernetstate} starting from a primary state $\ket{h}$. In fact, the fixed point $\tilde \eta_*$ depends only on the driving parameters $q\lambda/L$, $qT_0/L$ and $qT_1/L$ through~\eqref{eq:fixedpointetaff}, such that tuning these parameters enables one to generate as a steady state any given coherent state. Such a steady state engineering protocol is particularly relevant given the recently found quantum scar properties of the coherent states~\eqref{eq:cohernetstate}~\cite{Liska_2023}, and can be applied to any critical theory in (1+1)d.

In order to test our field-theoretical predictions, we now study a non-unitary driven critical model on the lattice, which can be seen as a particular case of~\eqref{latticemodeleq}. Concretely, we consider a free-fermionic lattice model taking the following form,
\begin{equation}
 H_{\rm lat} = -\frac{1}{2}\sum_{i=1}^L v_ic_{i+1}^{\dagger}c_i+h.c.,
\label{eq:latticemodel}
\end{equation}
where $c_i$ and $c_j^{\dagger}$ are fermionic operators satisfying the usual anticommutation relations, $\{c_i,c_j^{\dagger}\} = \delta_{ij}$, $\{c_i,c_j\}=\{c_i^{\dagger},c_j^{\dagger}\}=0$ \footnote{We assume for simplicity periodic boundary conditions, $c_{L+1}=c_1$, although the results can also be applied to open boundary conditions by using methods from boundary conformal field theory.}.
In the continuum, it is well described by the inhomogeneous CFT Hamiltonian~\eqref{eq:basicmodel} with $c=1$. We consider as the initial state the ground state at half-filling, and a two-step protocol matching the one used for the CFT evolution. This driven non-unitary lattice model was recently studied numerically~\cite{Wen_2024}, and we refer to App.~\ref{app:latcal} for details about its numerical implementation. The numerical results for both entanglement entropy and energy are compared with our CFT predictions in Fig.~\ref{fig:lattice}, showing remarkable agreement. Interestingly, the agreement improves as the measurement strength $\lambda$ increases; this is expected since increasing it constrains the system to stay at lower energies where CFT is a good approximation of the lattice dynamics. In that regard, the driven CFT gives a better prediction for the lattice dynamics in the non-unitary setting than in the unitary one; in the latter, Floquet heating eventually leads to a breakdown of the CFT description on the lattice at long times, while non-unitary evolution in the former helps to stabilize heating on the lattice.
We finally stress that the lattice model also saturates to a stroboscopic steady state at infinite times, and such a state is the lattice discretization of the Virasoro coherent state of the form~\eqref{eq:cohernetstate}, which should display quantum scar properties similarly to their CFT counterpart~\cite{Liska_2023}.

We close the section by mentioning possible generalizations of non-unitary Floquet drives beyond the $\mathfrak{sl}(2,\mathbb{R})$ algebra. A natural extension is to study general spatial deformations, which involve the full Virasoro algebra. The resulting time evolution preserves generalized Virasoro coherent states of the form
\begin{equation}
\ket{\{\xi_i\}_{i\geq 0},h}=e^{\sum_{i=0}^{\infty} (\xi_i L_i + \bar \xi_iL_{-i}) }\ket{h},
\label{eq:generalstate}
\end{equation}
and the time evolution of such states with a general Hamiltonian can again be encoded in a classical differential equation, generalizing the Riccati equation derived in App.~\ref{app:timeevolucoherent}.
In particular, such coherent states can again be seen as a displacement operator acting on primary fields~\cite{PhysRevB.103.224303,Liska_2023}.
Adding non-unitary time evolution also preserves the coherent states~\eqref{eq:generalstate}, leading to solvable generalizations of the non-unitary drive presented in this section. Although the precise study of this generalized setup and the computation of the entanglement entropy in generalized coherent states~\eqref{eq:generalstate} goes beyond the scope of this work, we expect a similar emergence of stroboscopic steady states to occur when the initial state is taken to be of the form~\eqref{eq:generalstate}.

\section{Beyond Floquet non-unitary circuits}
\label{sec:beyondfloquet}

In the context of driven CFTs in the absence of non-unitarity, different types of non-periodic dynamics have been extensively studied, given that the time evolution of observables can simply be encoded in products of SU(1,1) transformations. A first example consists of quasiperiodic drives, which follow a deterministic but never-repeating sequence. A canonical example of such evolution is generated by the Fibonacci sequence, and is called the Fibonacci quasiperiodic drive. In this case, it was shown that the heating-to-nonheating phase transition essentially disappears, only leaving a set of measure zero in parameter space that escapes heating~\cite{PhysRevResearch.2.033461, PhysRevResearch.3.023044}. More recently, it was found that other types of quasiperiodic sequences can lead to a heating-to-nonheating phase transition, similar to the periodic case~\cite{fang2025phasetransitionsquasiperiodicallydriven}. Furthermore, random drivings, which reduce to random walks in SL($2,\mathbb{R}$), have been shown to generically lead to heating, without any phase transition surviving~\cite{Wen_2022random}.  Beyond the $\mathfrak{sl}(2,\mathbb{R})$ subalgebra of the Virasoro algebra, we note that CFTs evolving under time-dependent noise coupled to the stress tensor were recently studied~\cite{PhysRevX.13.011043, Bernard_2020}. In this case, a steady state emerges at long times for a single noise realization, although averaging over the noise leads to heating with unbounded growth of entanglement.

Inspired by these recent developments, we now investigate the non-unitary dynamics of CFTs beyond the Floquet paradigm, using our formalism based on coherent states. Our goal is to understand whether a unique steady state emerges at long times, similarly to the periodic drive. To address this question, we first study an evolution that alternates quasiperiodically between unitary evolution and weak measurements, based on the Fibonacci sequence. Then, we consider a random switching between unitary evolution and weak measurements.

\subsection{Quasiperiodic non-unitary circuits}
\label{sec:fiboquasiperio}
\begin{figure}[htbp]
    \centering
        \includegraphics[width=0.48\textwidth]{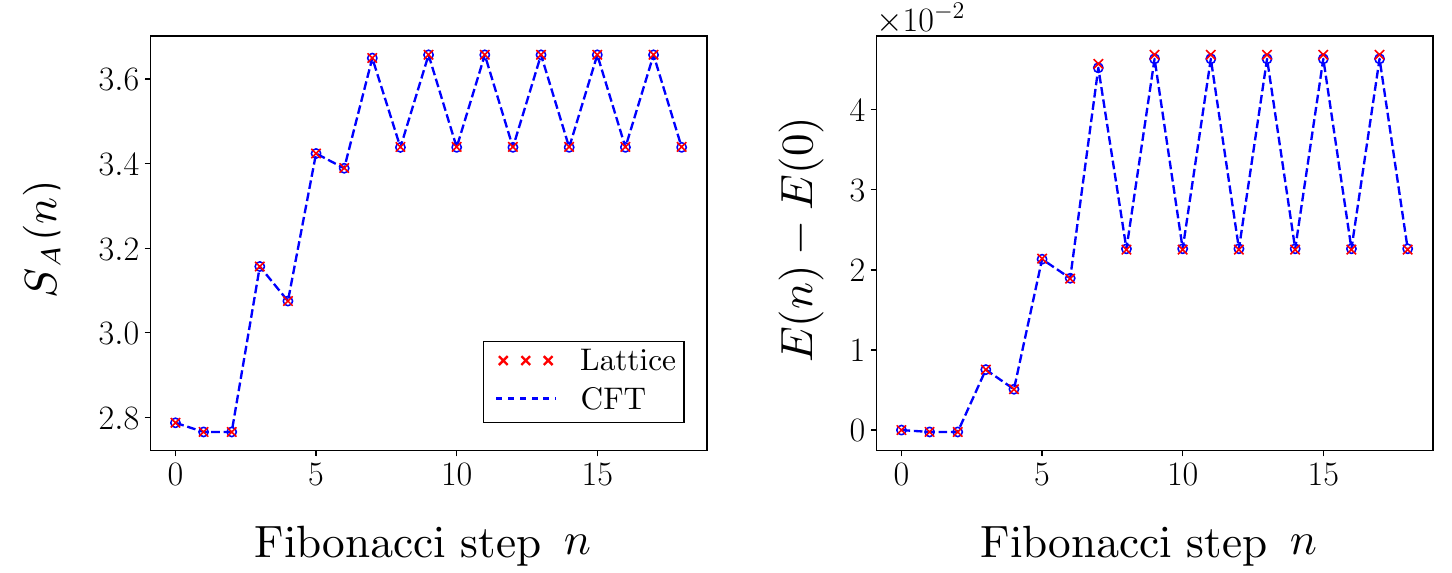}
    \caption{Time evolution of (a) the entanglement entropy $S_A(n)$ and (b) the total energy $E(n)$ after $n$ Fibonacci steps (equivalent to $F_{n+1}$ gates) of the quasiperiodic evolution given by~\eqref{eq:fibonaccirecursion}, for $\lambda=3.2$, $T/L=1/10$. We compare the CFT results (blue) and lattice calculations (red crosses), for a lattice system of size $L=800$ and periodic boundary conditions, starting from the ground state of the uniform chain. We observe that at large times, both the energy and entanglement entropy saturate to two distinct steady states depending on the parity of the Fibonacci time step $n$. We note that these emergent steady states are independent of the initial coherent state, i.e., do not depend on $\eta$.}
\label{fig:quasievolution}
\end{figure}

As a first step, we consider alternating between the non-unitary operator $U_0=e^{-\lambda H_0}$ and the unitary operator $U_1=e^{-i H_1 T}$ (with $H_1$ of the general form~\eqref{eq:defham}) in a deterministic but never-repeating way. This can be done using the Fibonacci quasiperiodic sequence
\begin{equation}
U_n = U_{n-2}U_{n-1},
\label{eq:fibonaccirecursion}
\end{equation}
such that at the Fibonacci step $n$, the string of operators, composed of $U_0$ and $U_1$, is of size $F_{n+1}$ (where $F_n$ denotes the $n$-th Fibonacci number). This sequence maps the initial state $\ket{\Psi_0}$ to 
\begin{equation}
\ket{\Psi_0} \mapsto \ket{\Psi_n}=\frac{U_n\ket{\Psi_0}}{||U_n\ket{\Psi_0}||},
\end{equation}
after $n$ Fibonacci steps.
Similarly to the periodic case, the dynamics of the coherent state can be traced back by looking at the iterated SL$(2,\mathbb{C})$ matrix $M_n$, following $M_n = M_{n-2}M_{n-1}$, with $M_0$ and $M_1$ defined as in~\eqref{eq:NonUnitary_M}. In the unitary setting, this problem could be mapped to quasicrystals using the so-called Fibonacci trace map~\cite{PhysRevResearch.2.033461, PhysRevResearch.3.023044}, which leads to a fractal phase diagram with the non-heating region forming a Cantor set. However, as long as the measurement strength is nonzero, $\lambda\neq 0$, we find that such a fractal structure is absent in the time evolution of energy or entanglement entropy; see App.~\ref{app:quasiper} for an analytical argument based on mappings to quasicrystals.
Instead of infinitely heating up, the system reaches a steady state that is independent of the initial (coherent) state $\ket{\Psi_0}$, similarly to the periodic case detailed in Sec.~\ref{sec:periodicnonunitary21}. Nevertheless, an intriguing feature arises which does not take place in the periodic circuit: Two distinct steady states emerge for even and odd Fibonacci times, i.e.,
\begin{equation}
\lim_{n\rightarrow\infty}\ket{\Psi_{2n}} = \ket{\Psi_{*,1}}, \quad \lim_{n\rightarrow\infty}\ket{\Psi_{2n+1}} = \ket{\Psi_{*,2}},
\end{equation}
where both $\ket{\Psi_{*,1}}$ and $\ket{\Psi_{*,2}}$ are determined numerically; see e.g. Fig.~\ref{fig:quasievolution}.
Therefore, the stable fixed points in the unit disk are distinct for even and odd Fibonacci times $n$. This feature is reminiscent of the so-called time quasicrystals~\cite{PhysRevLett.120.070602}, where similar persistent oscillations were found as a function of the Fibonacci time steps $n$ (although in that case, the persistent oscillations were found to be period-3). We note that similar behavior has also recently been observed in non-unitary fermionic Gaussian circuits~\cite{lapierreryu2025}. As a direct consequence, both energy and entanglement entropy oscillate with period two between the two steady states, as shown in Fig.~\ref{fig:quasievolution}(a-b). Moreover, the lattice calculations, based on the free fermionic model~\eqref{eq:latticemodel}, also show the emergence of two steady states which agree with the CFT predictions, even after relatively large gate numbers, of the order of $F_{21}\sim 10^4$.

Finally, we note that generalizations of the Fibonacci quasiperiodic drive were recently studied as minimal models of complete Hilbert space ergodicity~\cite{PhysRevLett.131.250401}. These generalized sequences are defined as $U_n = U_{n-2}^{m_1}U_{n-1}^{m_2}$, for positive integers $m_1$ and $m_2$ (we recall that $m_1=m_2=1$ generates the usual Fibonacci word). We observe the occurrence of the even-odd steady states even in these generalized driving sequences, suggesting it to be a robust feature inherited from the Fibonacci sequence.

It will be interesting to further generalize this to other quasiperiodic sequences, such as Thue-Morse~\cite{PhysRevX.7.031034}, multipolar drivings~\cite{PhysRevLett.126.040601},  or Aubry-André~\cite{PhysRevResearch.3.023044} type models. In particular, it is an open question whether the phase transition recently discovered in an Aubry-André-type quasiperiodically driven CFT~\cite{fang2025phasetransitionsquasiperiodicallydriven} may still arise when adding non-unitary terms.

\subsection{Random non-unitary circuits}

We now extend our analysis to random non-unitary time evolution. Let us consider for simplicity an alternation between unitary evolution $e^{-i T_0 H_1}e^{-i T_1 H_1}$ and weak measurement $e^{-\lambda H_0}$. At the level of coherent states, the random evolution reduces to a random product of the matrices $M_1$ and $M_0$ defined as~\eqref{eq:NonUnitary_M}. Therefore, mathematically, the setup maps to a random walk on the hyperbolic disk. The asymptotic behavior of the random products of  SL$(2,\mathbb{C})$ matrices is known to be classified by Furstenberg's theorem~\cite{10.1214/aoms/1177705909}. The theorem states that the Lyapunov exponent of such a random product is strictly positive under the assumption that the smallest subgroup of SL$(2,\mathbb{C})$ containing the random product is non-compact and strongly irreducible. More precisely, we define the Lyapunov exponent as
\begin{equation}
\lambda_{\text{L}} =\lim_{n\rightarrow\infty} \frac{1}{n}\log||\Pi_n||, 
\end{equation}
where $\Pi_n$ denotes a random sequence of length $n$, e.g., $\Pi_n=M_0M_0M_1M_0M_1M_1...$ . In the unitary case, it has been shown recently that $\lambda_{\text{L}}>0$ implies that the random walk on $\mathbb{D}$ converges towards $\partial \mathbb{D}$. Physically, this implies that the averaged entanglement entropy (averaged energy) grows linearly (exponentially), i.e., that heating is unavoidable, apart from exceptional points of measure zero in parameter space that violate the assumptions of Furstenberg's theorem and have zero Lyapunov exponent, $\lambda_{\text{L}}=0$~\cite{Wen_2022random}. One of such exceptional points corresponds to the random walk generated by $M_0$ and $M_1$ under the constraint $\text{Tr}(M_0) = \text{Tr}(M_1)=0$. We now apply these ideas to our non-unitary drive, where the random walk takes place on SL$(2,\mathbb{C})$ instead of SL$(2,\mathbb{R})$. In our context, $\lambda_{\text{L}}>0$ implies that the time-evolved state at large times is independent of the initial coherent state. In particular, we note that as long as measurement strength is nonzero, $\lambda\neq0$, the trace of the associated SL$(2,\mathbb{C})$ matrix is nonzero, which implies that the exceptional points discovered in the unitary setting in~\cite{Wen_2022random} do not exist anymore in the non-unitary generalization. This is consistent with the inexistence of an oscillatory phase for the periodic non-unitary drive, as discussed in Sec.~\ref{sec:periodicnonunitary21}.

\begin{figure}[htbp]
    \centering
        \includegraphics[width=0.48\textwidth]{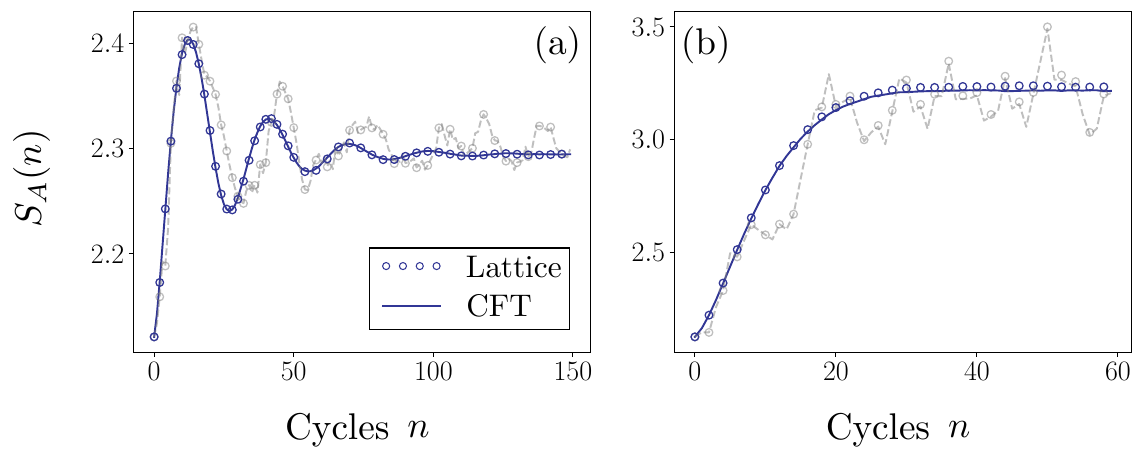}
    \caption{Time evolution of the entanglement entropy $S_A(n)$ for a random two-step non-unitary drive. With probability $p= \frac{1}{2}$ we drive the system with $e^{-iT_0 H_0}e^{-iT_1 H_1}$, while with probability $1-p = \frac{1}{2}$ we perform a weak measurement of the form $e^{- \lambda H_0}$. As before, we choose parameters $(\sigma_0, \sigma_+, \sigma_-) = (1, 0, 0)$ in~\eqref{eq:defham} for $H_0$, and $(0, 1, 0)$ for $H_1$. For plot (a), the value of $\lambda$ is $0.4$, while for plot (b) the value of $\lambda$ is $0.2$. For each plot, we show an example of a single realization (dashed gray line), and an average over 4000 realizations (solid blue line), and compare them with free-fermionic lattice calculations (hollow dots) with model~\eqref{eq:latticemodel}, for a system of size $L=200$ and periodic boundary conditions, starting from the ground state of the uniform chain. The two distinct behaviors observed for the Floquet evolution in Fig.~\ref{fig:lattice}, corresponding to damped oscillations ($T_0/L=1/50$, $T_1/L=1/100$) and growth and then saturation to the steady state ($T_0/L=1/100$, $T_1/L=1/50$), are still present upon averaging over randomness.}
\label{fig:random_evolution}
\end{figure}

As a concrete protocol to test the above general considerations, the alternation is implemented considering a random series extrapolated from a binomial distribution with probability $p$ of drawing the unitary evolution, and $(1-p)$ of drawing the non-unitary one. A particularly natural choice , which allows a comparison with the Floquet evolution, is $p = \frac{1}{2}$, even though such restriction is not necessary and the features discussed below remain for arbitrary $p$. The physical interpretation of this protocol could be that of a random sequence of weak measurements interspersing the unitary driven CFT evolution. The results for the half-size entanglement entropy are shown in Fig.~\ref{fig:random_evolution}. Single random realizations are depicted as dashed gray lines, and never reach a steady state. On the other hand, averaging over random sequences leads to an emergent steady state, whose value turns out to be close to the Floquet one, as one would expect for $p=\frac{1}{2}$. Additionally, the timescales of the averaged evolution correspond to the ones of the Floquet case. In particular, the period of the oscillations in Fig.~\ref{fig:random_evolution}(a) and the growth-saturation timescale in Fig.~\ref{fig:random_evolution}(b) correspond exactly to the Floquet ones.
As a final qualitative comment, we notice that the amplitudes of the random fluctuations for a typical random realization of the would-be heating phase (Fig.~\ref{fig:random_evolution}(b)) are much higher than the amplitudes of the nonheating phase (Fig.~\ref{fig:random_evolution}(a)). This feature is inherited by the respective unitary evolution, which in the absence of non-unitarity leads to linear growth in one case, and bounded oscillations in the other.

\section{Purification phase transitions}
\label{sec:phasetransitions}

\begin{figure}[thbp]
    \centering
        \includegraphics[width=0.85\columnwidth]{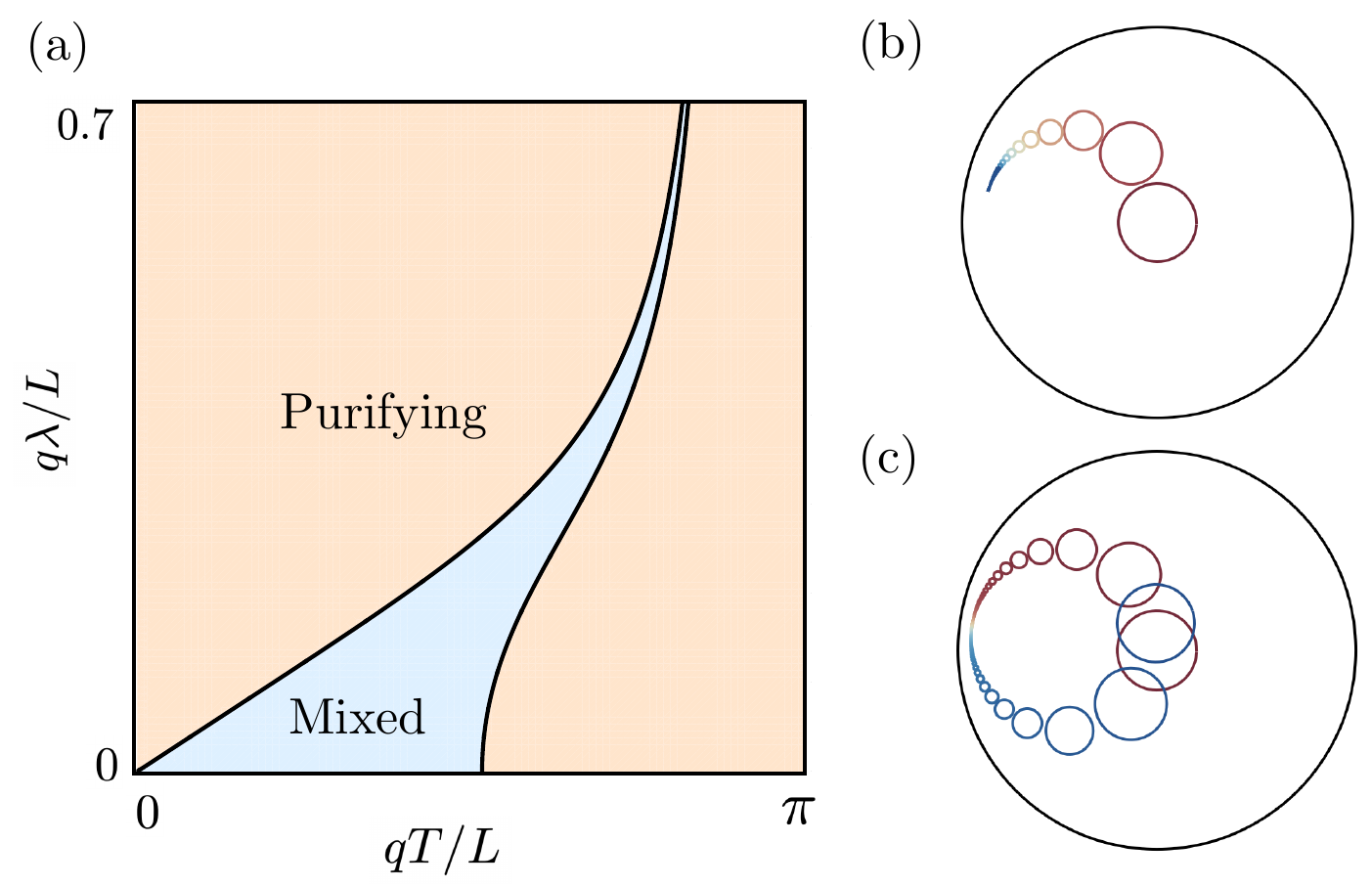}
    \caption{(a) Phase diagram of the purification transition for a Floquet circuit of the form~\eqref{eq:non-unitary_evolution_positive_and_negative} in the case $\lambda_1 = \lambda_2 \equiv \lambda$. We choose parameters $(\sigma_0, \sigma_+, \sigma_-) = (0.3, 0.5, 0)$ in~\eqref{eq:defham} for $H_1$, and $(0.5, 0.3, 0)$ for $H_2$, such that the phase boundary is given by~\eqref{eq:phaseboundary}. Two phases emerge: a purifying phase where an initial mixture of coherent states (e.g. centered around the origin of $\mathbb{D}$, corresponding to the density matrix~\eqref{eq:densitymatrix}) evolves towards a pure state~\eqref{eq:evolvepure}, and a mixed phase where the initial mixture of coherent states fails to converge to a pure state even at infinite times due to the absence of a fixed point in the one-cycle map~\eqref{eq:onecyclemappurifica}. (b-c) Illustration of the two types of evolutions for an initial circular configuration centered around the origin, where stroboscopic time evolution goes from red to blue, and maps circles to circles.}
\label{fig:phase_diagram}
\end{figure}

In the previous sections, we have demonstrated that the competition between a unitary drive and postselected weak measurements of the form $e^{- \lambda H_0}$ systematically leads to the stabilization of steady states that are independent of the initial state, even for aperiodic and random drivings. 
Therefore, the heating-to-nonheating phase transition is washed away as long as $\lambda>0$.
We now consider a more general class of non-unitary operators, such that a new type of transition, with no counterpart in the unitary setting, emerges. We will frame this phase transition in terms of a purification transition~\cite{PhysRevX.10.041020, Fidkowski_2021, PhysRevLett.126.060501} for an initially mixed state built out of coherent states. 
Within the evolution, we will also consider non-unitary operators of the form $e^{\lambda H_0}$, representing yet another type of weak measurement, as discussed in App.~\ref{app:postselection}.
We design a periodic non-unitary circuit where the two postselected weak measurements are alternated with a unitary evolution of the form introduced in Section~\ref{sec:Unitary_Floquet_drives}. Concretely, we consider a one-step drive of our evolution of the form
\begin{equation}
    U_{\text{F}} = e^{- \lambda_1 H_0} e^{-i T H_2} e^{\lambda_2 H_0} e^{-i T H_1} ,
    \label{eq:non-unitary_evolution_positive_and_negative}
\end{equation}
where $H_1$ and $H_2$ are of the form~\eqref{eq:defham}.
As discussed previously around \eqref{eq:NonUnitary_M}, this can be written in terms of an SL$(2, \mathbb C)$ transformation that acts on a coherent state $\ket{\eta,h}$ through
\begin{equation}
\label{eq:onecyclemappurifica}
    M_{\text{F}} = M_0(\lambda_1) M_2 M_0^{-1} (\lambda_2) M_1\, ,
\end{equation}
where we have used the same notation as in \eqref{eq:NonUnitary_M}, except for 
\begin{equation}
    M_{0}(\lambda) \equiv \begin{pmatrix}
    e^{- \frac{\pi\lambda q}{L}} & 0 \\ 0 & e^{ \frac{\pi \lambda q}{L}}
    \end{pmatrix} .
\end{equation}
It is straightforward to show that for $\lambda_1 > \lambda_2$ the system always reaches a steady state, as in Sec.~\ref{sec:periodicnonunitary21}, while for $\lambda_2 >\lambda_1$ the energy always diverges in a finite amount of time, which is therefore an unphysical scenario that we discard. On the other hand, intriguing physics emerges in the special case $\lambda_1  = \lambda_2 \equiv \lambda$, where the measurement strength for both types of measurements is equal. 
In this case, the trace of the one-cycle SL$(2,\mathbb{C})$ transformation $M_{\text{F}}$ takes the form
\begin{equation}
\Tr(M_{\text{F}})=2\text{Re}(a_2a_1)+e^{-\lambda}\bar b_2 b_1+e^{\lambda} \bar b_1 b_2.
\end{equation}
In particular, by imposing $\text{arg}(b_1)=\text{arg}(b_2)$, the trace is real, $\text{Im}(\Tr(M_{\text{F}}))=0$.
Importantly, for some choices of parameters, there can be a transition from a hyperbolic transformation to an elliptic transformation, the latter corresponding to $|\text{Re}(\Tr(M_{\text{F}}))|<2$ and $\text{Im}(\Tr(M_{\text{F}}))=0$. This implies that there can be a transition between a case where there is no stable fixed point within $\mathbb{D}$, to a case where such a fixed point emerges. Such an example can be obtained by taking $(\sigma_0,\sigma_+,0)$ for $H_1$ and $(\sigma_+,\sigma_0,0)$ for $H_2$ in~\eqref{eq:defham}. In that case, the phase boundary is given by
\begin{flalign}
\label{eq:phaseboundary}
    &1= \bigg|\cos(\sqrt{C}\frac{\pi q T}{L})\cosh(\sqrt{C}\frac{\pi qT}{L}) \notag+\frac{\sigma_0\sigma_+}{C} \times \\
    & \times\left[\cosh(\frac{\pi q\lambda}{L})-1\right]\sin(\sqrt{C}\frac{\pi qT}{L})\sinh(\sqrt{C}\frac{\pi qT}{L})\bigg|,
\end{flalign}
with $C=\sigma_0^2-\sigma_+^2$.
The resulting phase diagram is shown for the above example in Fig.~\ref{fig:phase_diagram}(a) with $\sigma_0=0.3$ and $\sigma_+=0.5$.
This transition in the fixed point of the Floquet map gives rise to a purification phase transition when the initial state is mixed. Indeed, the initial density matrix may or may not purify at infinite times depending on whether there is a stable fixed point in the Floquet map. We investigate how the transition arises when the initial state is a mixture of coherent states within a single Verma module. Concretely, we consider the following initial mixed state, combining coherent states of equal energy,
\begin{align}
\label{eq:densitymatrix}
    \rho_0 = &  \oint \frac{\dd\varphi}{2 \pi} \ketbra{|\eta| e^{i \varphi}, h}{|\eta| e^{i \varphi}, h} \nonumber \\
    = & (1 - |\eta|^2)^{2h_q} \sum_{n = 0}^\infty |\eta|^{2n} \frac{\Gamma(2 h_q + n)}{\Gamma(n+1) \Gamma(2h_q) } \ketbra{h,q,n}{h,q,n},
\end{align}
where in the second equality we have used the expression of coherent states ~\eqref{eq:Coherent_state_explicit_form} in terms of $\ket{h,q,n}$, the (normalized) descendant states $L_{-q}^n \ket{h}$.
Given that the non-unitary evolution~\eqref{eq:non-unitary_evolution_positive_and_negative} is not trace preserving, we define the time evolution of the density matrix to be
\begin{equation}
    \tilde \rho_{n} = \frac{U_\text{F} \tilde \rho_{n-1} U_{\text{F}}^{\dagger}}{\text{Tr}(U_\text{F} \tilde \rho_{n-1} U_{\text{F}}^{\dagger})},
\end{equation}
similarly to~\eqref{eq:coherentstatepresr} for pure states. The long-time dynamics of the density matrix is dictated by whether there is a stable fixed point or not: if a stable fixed point $\tilde \eta_*$ exists within $\mathbb{D}$, then each individual coherent state $||\eta|e^{i\varphi},h\rangle$ in~\eqref{eq:densitymatrix} converges to $\ket{\tilde \eta_*,h}$ as $n\rightarrow\infty$. As a consequence, the density matrix $\rho_0$ maps at infinite time to
\begin{equation}
\lim_{n\rightarrow\infty}\tilde \rho_n = \ketbra{\tilde \eta_*,h},
\label{eq:evolvepure}
\end{equation}
i.e., a pure state described by a single coherent state, as illustrated in Fig.~\ref{fig:phase_diagram}(b). On the other hand, if the one-cycle transformation is elliptic, there is no stable fixed point within the unit disk and each individual coherent state $||\eta|e^{i\varphi},h\rangle$ in~\eqref{eq:densitymatrix} oscillates as a function of $n$ --- therefore $\rho_0$ does not purify even at infinite times, as shown in Fig.~\ref{fig:phase_diagram}(c). We note that similar purification phase transitions were recently found in non-unitary free fermionic Floquet circuits, where the time evolution of fermionic coherent states was similarly encoded in SL$(2,\mathbb{C})$ transformations~\cite{Granet_2023} (see~\cite{zhang2023antiunitarysymmetrybreakinghierarchy} for a recent work in Floquet interacting models).  In these cases, the transition has been shown to emerge due to the breaking of PT symmetry in the Floquet Hamiltonian $H_{\text{F}}$, defined as $U_{\text{F}} = e^{-H_{\text{F}}}$. It will be an interesting question to study whether the CFT purification phase transition similarly arises due to PT symmetry breaking.

\begin{figure}[thbp]
    \centering
        \includegraphics[width=\columnwidth]{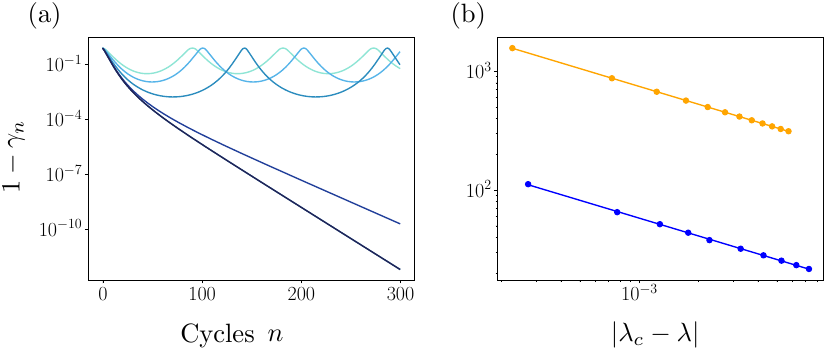}
    \caption{(a) Time evolution of the purity $\gamma_n$ after $n$ Floquet cycles (with the same parameters for the Floquet operator as in Fig.~\ref{fig:phase_diagram}) for an initial density matrix given by~\eqref{eq:densitymatrix} (with $|\eta|=0.7$, $q = 1$ and $h = 1$), and $T/L = 0.2$. Values of the measurement strength $2 \pi \lambda/L$ range within $\{0.05, 0.08, 0.12, 0.16, 0.17\}$ (light to dark blue), with transition at approximately $0.15$. In the mixed phase, the purity oscillates in time with a parameter-dependent periodicity, and thus never converges to one. In the purifying phase, the purity exponentially converges to one. (b) Critical behavior near the purification phase transition, at the critical measurement rate $\lambda_c$. The orange fit (upper curve) corresponds to the divergence of the periodicity of the purity in the mixed phase. The blue fit (lower curve) corresponds to the divergence of the inverse of the purification rate in the purifying phase.
    In both cases, we numerically extract the critical exponent to be approximately $0.5\pm0.002$. }
\label{fig:phase_diagram2}
\end{figure}

In order to characterize the purification transition, we compute the stroboscopic time evolution of the purity of $\tilde \rho_n$, defined as $\gamma_n = \Tr(\tilde \rho_n^2)$. We recall that while for the unitary time evolution discussed in Sec.~\ref{sec:Unitary_Floquet_drives} the purity does not evolve in time, non-unitary evolution necessarily changes the purity.
Throughout the evolution, we need to be careful that at each stroboscopic time $n$ the state is well-defined, i.e., no points exceed the boundary of the unit disk, as it would lead to unphysical results.
We find that for a measurement strength $\lambda$ sufficiently low the procedure is well-defined.  Fixing $\lambda$ below the allowed threshold, the system displays a purification phase transition as a function of the driving parameters $(\frac{qT}{L}, \frac{q\lambda}{L})$, as shown in Fig.~\ref{fig:phase_diagram}. In particular, we observe that the mixed phase eventually vanishes in the limit of large $\frac{q\lambda}{L}$ or large $\frac{qT}{L}$, while the purifying phase appears to be the generic scenario --- this is consistent with the pure state case studied in Sec.~\ref{sec:periodicnonunitary21}, where the emergence of a unique steady state at late time is the generic scenario. 
In the mixed phase, the purity oscillates as a function of time, as shown in Fig.~\ref{fig:phase_diagram2}(a). Importantly, the transition to the purifying phase can be characterized by the periodicity of the oscillations of the purity. We find that such periodicity scales as $|\lambda_c-\lambda|^{-\nu}$, where $\lambda_c$ is the critical measurement strength. We numerically extract the critical exponent $\nu=0.5\pm0.002$ on Fig.~\ref{fig:phase_diagram2}(b). Alternatively, the same critical exponent can be extracted from the purification rate by approaching the mixed phase from the purifying phase.
This value of the critical exponent is universal in the sense that it is the same for any (1+1)d CFT subject to our non-unitary evolution. We note that $\nu=\frac{1}{2}$ is expected on general grounds, given that the dynamics is constrained by the SL$(2,\mathbb{C})$ group. In fact, it appears in other contexts where related transitions from elliptic to hyperbolic orbits occur such as heating-to-nonheating transitions~\cite{wen2018floquetconformalfieldtheory}, and even in the study of light-heavy backgrounds and black hole formation in holography~\cite{Anous_2016, Anous_2019}.
Nevertheless, this exponent differs from the exponents found in random quantum circuits~\cite{PhysRevX.10.041020}, as well as the ones recently found in non-unitary integrable Floquet circuits~\cite{zhang2023antiunitarysymmetrybreakinghierarchy}.

Our analysis was so far restricted to mixed states defined from a specific subspace of a given Verma module, generated by states of the form $\{L_{-q}^n\ket{h}\}_{n\geq 0}$.
It is, however, an open question whether the purification transition still appears when considering mixed states obtained from general descendant states, in which case our approach based on $\mathfrak{sl}(2,\mathbb{R})$ coherent state does not apply, and one needs to invoke the full Virasoro algebra.
We expect that the phase transition is altered when the initial density matrix mixes states from different Verma modules. In fact, our non-unitary drive is generated by conformal transformations, and thus does not mix different Verma modules. This directly implies that there is a different steady state for each Verma module, and thus the infinite time density matrix remains partially mixed. 

Finally, we conclude this section by drawing some connections to measurement-induced phase transitions. As discussed in the context of monitored random quantum circuits in~\cite{PhysRevX.10.041020}, it is expected that the purification phase transition for mixed initial states happens concurrently with an entanglement phase transition for pure initial states. When the initial state $\ket{\Psi(0)}$ is an arbitrary superposition of coherent states $\ket{\eta,h}$ within a single Verma module, i.e.,
\begin{equation}
\ket{\Psi(0)} = \sum_{i=1}^{\infty}\theta_i|\eta_i,h\rangle,
\end{equation}
it is clear that the time-evolved state $\ket{\Psi(n)}$ maps at infinite times to a single coherent state $\ket{\tilde \eta_*,h}$ in the purifying phase. As was discussed in Sec.~\ref{sec:Setup}, if we assume that $h=0$, the entanglement entropy of a coherent state $S_A(\eta,0)$ follows a logarithmic violation of the area law, similar to the ground state $\ket{0}$. On the other hand, in the mixed phase, the state $\ket{\Psi(n)}$ remains an arbitrary superposition of coherent states at any time, which may thus follow e.g. volume law entanglement. Thus, we expect an entanglement phase transition to similarly arise for pure initial states. We leave the precise study of such transitions in CFT and in critical lattice models for future works.

\section{Discussion}
\label{sec:Conclusion}

In this work, we have studied the dynamics of CFTs subjected to both unitary dynamics and postselected weak measurements. In doing so, we have uncovered universal dynamics for the entanglement entropy and energy, and showed that a multitude of rich features appears: a unique steady state emerges for Floquet drives, a single {\it averaged} steady state emerges for random driving, and a discrete set of steady states emerges for quasiperiodic drives, {\it e.g.} two for the case of the Fibonacci circuit.

Moreover, we uncovered intriguing physics as we allowed for the competition between different types of non-unitary operators: when starting from a class of initial mixed states, we have identified purification phase transitions, similar to those discovered in the context of measurement-induced phase transitions in random non-unitary circuits.
The transition separates a fully mixed phase that does not purify at infinite times, and a purifying phase following exponential purification to a single coherent state. The nature of the phase transition could be understood in simple geometric terms as the emergence of a fixed point of the Floquet map that encodes the evolution of the density matrix.

Our work thus paves the way towards the exploration of universal dynamics of critical systems subjected to both nonequilibrium unitary evolution and measurements. 
\noindent In the following, we outline several future research directions.

\textit{Stochastic measurements} --- Throughout this paper, we have considered imaginary evolution with the CFT Hamiltonian as a model of a weakly measured quantum critical system with postselection, as motivated by recent works on the driven non-unitary Ising chain~\cite{PhysRevResearch.6.013131}. It is tempting to extend our results to continuous monitoring, where the time evolution is described by the stochastic Schrödinger equation~\cite{Wiseman:2009rda}. Given the underlying algebraic structure of our problem, and the fact that coherent states are left invariant under the action of Virasoro generators, we may expect that the growth of e.g. entanglement entropy under stochastic measurements can be studied analytically. 

\textit{Generalizations to the Virasoro algebra} --- A general class of non-unitary circuits that can be studied using coherent state techniques in CFT are those involving arbitrary spatial deformations of the stress tensor, i.e. involving the full Virasoro algebra. In fact, such evolution preserves the generalized coherent states~\eqref{eq:generalstate}. While the time evolution of such states cannot simply be parametrized as a classical motion in the hyperbolic disk, their dynamics can be classified through the co-adjoint orbits of the Virasoro group~\cite{Witten:1987ty}. The resulting circuit is a non-unitary extension of the so-called Virasoro circuit~\cite{PhysRevLett.122.231302, deBoer:2023lrd, Flory_2020}, and should naturally lead to multiple transitions between purifying and mixed phases, enriching the purification transition found in Sec.~\ref{sec:phasetransitions}.

\textit{Connection to Krylov complexity} --- Over the past years, Krylov complexity has emerged as a central notion in both quantum information~\cite{Parker:2018yvk} and high energy theory~\cite{Rabinovici:2023yex}. In particular, several recent works have focused on the dynamics of Krylov complexity for Hamiltonians constrained by the $\mathfrak{sl}(2,\mathbb{R})$ algebra~\cite{caputa2021geometrykrylovcomplexity, malvimat2024krylovcomplexity2dcfts}, and other recent developments applied the notion of Krylov complexity to unitary Floquet circuits~\cite{PhysRevB.111.014309}. It would be very interesting to understand the purification transition of Sec.~\ref{sec:phasetransitions} in terms of Krylov complexity, which has proved to be a sensitive probe of dynamical aspects of quantum systems.

\textit{Holography} --- It would be desirable to reinterpret the results derived here within the context of the AdS/CFT correspondence. Recent works have focused on studying the holographic dual of coherent states~\cite{Caputa_2023}, and of driven CFTs through the lens of dynamical event horizons~\cite{goto2021nonequilibratingblackholeinhomogeneous, deBoer:2023lrd, miyata2024hawkingpageentanglementphasetransition, li2025holography2dinhomogeneouslydeformed}, operator reconstruction through modular flow~\cite{jiang2024newhorizonsinhomogeneousquenches}, and from the point of view of end-of-the-world branes~\cite{10.21468/SciPostPhys.15.5.202}. Generalizing these works to the present context is rather natural, as our non-unitary evolution $e^{-\lambda H_0}$ can be implemented by considering an evolution over imaginary times on the boundary, which turns into a complex geometry in the bulk. Interestingly, the dynamical event horizon that appears in the unitary heating phase is destroyed by the measurements, which localize the state into a late-time coherent steady state. From the point of view of the AdS/CFT correspondence this is a rather peculiar phenomenon, showing a concrete example where boundary measurements can change the causal structure of the bulk, which is expected to be related to the entanglement structure of the boundary state \cite{VanRaamsdonk:2010pw, Maldacena:2013xja}, and is certainly affected by measurements.
On the other hand, a bulk interpretation of the purification phase transition is presumably more ambitious to achieve, as both the initial `squeezed' mixed state~\eqref{eq:densitymatrix} and the non-unitary evolution $e^{\lambda H_0}$ do not seem to have a clear bulk interpretation. 
More broadly speaking, the simplicity of the techniques presented here will allow a deeper understanding of the interplay between geometry, entanglement and measurements within holography, exploiting the rich phenomenology of non-equilibrium physics.


\begin{acknowledgments}

We thank Diego Liska, Jiri Minář, Adam Nahum and Marco Schirò for inspiring discussions. 
B.L. thanks Valerio Pagni, Per Moosavi, and Ramasubramanian Chitra for previous collaborations related to non-Hermitian CFTs. This research is supported in part by the Fonds National Suisse de la Recherche Scientifique (Schweizerischer Nationalfonds zur Förderung der wissenschaftlichen Forschung) through the Project Grant 200021\_215300 and the NCCR51NF40-141869 The Mathematics of Physics (SwissMAP). J.S.\ thanks Harvard University for support through a Bershadsky Distinguished Visiting Fellow award. P.P.\ thanks Princeton University for hospitality during which part of this work was carried out.
B.L.\ acknowledges financial support from the Swiss National Science Foundation (Postdoc.Mobility Grant No.~214461).
S.R.\ is supported
by the National Science Foundation under Award No.\ 
DMR-2409412. 
\end{acknowledgments}


\appendix

\section{Non-unitary evolution from weak measurements}
\label{app:postselection}

In this appendix we show that the non-unitary evolution considered in this work can be realized by coupling the system with a single-qubit ancilla, and postselecting projective measurement on the ancilla. The general idea is to employ the following identity which involves Pauli matrices of two-level systems, 
\begin{equation}
    e^{- i \mathcal A \otimes \sigma_a^y} = \cos(\mathcal A) \otimes \mathbbm{1}_a - i \sin(\mathcal A) \otimes \sigma_y^a \ ,
    \label{eq:pauli_identity}
\end{equation}
where the trigonometric matrices as functions of $\mathcal A$ are defined through a power series. Therefore, defining 
\begin{equation}
    \cos(\mathcal A) = e^{- \lambda (H - E_0)} \ ,
    \label{eq:postselection_evolution_definition_A}
\end{equation}
we find that our imaginary time evolution can be seen as a postselected measurement on the ancilla, after evolving both the system and the ancilla as~\eqref{eq:pauli_identity} and then projecting
\begin{equation}
    e^{- \lambda (H-E_0)} = {}_a \! \bra{0} e^{- i \mathcal A \otimes \sigma_a^y} \ket{0}_a \ .
\end{equation}
The shift in energy in the Hamiltonian vanishes when normalizing the state after the postselection. Additionally, it was demonstrated in~\cite{PRXQuantum.2.010342} that such imaginary evolution through an ancilla can be implemented in quantum simulators for various lattice models, such as the transverse field Ising model at the critical point. 

A similar approach can also be followed for an evolution of the form $e^{\lambda H}$ introduced in Sec.~\ref{sec:phasetransitions}. In this case, changing the sign of~\eqref{eq:postselection_evolution_definition_A} is not a feasible strategy, as all the eigenvalues of $e^{\lambda (H-E_0)}$ are larger than one, and thus $\mathcal A$ would not be Hermitian. In order to circumvent this issue, one can truncate the Hilbert space of $H$, and consider the truncated Hamiltonian
\begin{equation}
    H = \sum_{n = 0}^\infty E_n \ketbra{E_n}{E_n} \; \to \;  \tilde H = \sum_{n = 0}^N E_n \ketbra{E_n}{E_n}
\end{equation}
with $E_N \equiv \mathcal E$ being a very high cutoff energy. Then, we consider
\begin{equation}
    \cos \big( \tilde{\mathcal A} \big) = e^{\lambda (\tilde H - \mathcal E)} \ .
    \label{eq:postselection_evolution_definition_A_amplification}
\end{equation}
In this way all eigenvalues of $e^{\lambda (\tilde H - \mathcal E)}$ are smaller than one, and the Hermitian matrix $\tilde{\mathcal A}$ is well defined. Physically, we are considering a postselection that favors high energy states rather than low energy states. Clearly, if the spectrum is unbounded from above, we have to put a cutoff $\mathcal E$. This procedure works only if the dynamical evolution of the state involves mainly eigenstates with energy less than $\mathcal E$. In this case we have
\begin{equation}
    e^{\lambda (\tilde H- \mathcal E)} = {}_a \! \bra{0} e^{- i \tilde{\mathcal A} \otimes \sigma_a^y} \ket{0}_a \ ,
\end{equation}
is a good approximation to the amplification considered in Sec.~\ref{sec:phasetransitions}. To show it precisely, one can estimate
\begin{equation}
    || e^{\lambda H}\ket{\eta, h} - e^{\lambda \tilde H}\ket{\eta, h} ||^2 \sim |\eta|^{2 \mathcal E} \ ,
\end{equation}
which is a small deviation since $\mathcal E \gg 1$ and $|\eta| < 1$.

\section{Time evolution of coherent states}
\label{app:timeevolucoherent}

The aim of this appendix is to derive the formulas for the unitary and non-unitary evolution of coherent states~\eqref{eq:cohernetstate}, given by~\eqref{eq:transformationlaw} and~\eqref{eq:coherentstatepresr}, respectively. 

Concretely, we want to compute
\begin{equation}
e^{-i\frac{2\pi t}{L}\left[\sigma_0 L_0 + \frac{\sigma_+}{2}(L_q+L_{-q})+\frac{\sigma_-}{2i}(L_{q}-L_{-q})\right]}\ket{\eta,h}.
\end{equation}
In order to make progress, we introduce rescaled Virasoro generators that satisfy an $\mathfrak{sl}(2,\mathbb{R})$ algebra,
\begin{equation}
\label{mm}
\tilde L_0\equiv
\frac{1}{q}\Big(L_0+\frac{c}{24}(q^2-1)\Big)  ,
\quad
\tilde L_{\pm1}\equiv
\frac{1}{q}L_{\pm q} .
\end{equation}
Then, we need to evaluate, up to an overall phase,
\begin{equation}
e^{-\frac{2\pi i t q}{L} \left[\sigma_0\tilde{L}_0+ \frac{1}{2}(\sigma_+-i\sigma_-)\tilde{L}_{1}+ \frac{1}{2}(\sigma_++i\sigma_-)\tilde{L}_{-1} \right]}\ket{\eta, h}.
\end{equation}
As is well-known in the context of quantum optics, the $\mathfrak{sl}(2,\mathbb{R})$ Perelomov coherent states follow a classical equation of motion under such time evolution~\cite{PhysRevA.31.2721, PhysRevA.39.5717}. The equation is given for the unit disk coordinate $\eta$ as
\begin{equation}
\dot{\eta}=\{\eta,\langle \eta|H|\eta\rangle\},
\label{eq:classicalpoiusson}
\end{equation}
where $\{A,B\}$ is a generalized Poisson bracket of the form
\begin{equation}
\{A,B\}=\frac{(1-|\eta|^2)^2}{2ih_q}\left(\frac{\partial A}{\partial \eta}\frac{\partial B}{\partial \bar{\eta}}-\frac{\partial A}{\partial \bar{\eta}}\frac{\partial B}{\partial \eta}\right).
\end{equation}
In order to simplify this equation, we evaluate the classical Hamiltonian
$\mathcal{H}\equiv\langle \eta,h|H|\eta,h\rangle$,
\begin{multline}
    \mathcal{H} = \frac{2\pi q}{L}\Bigl[\sigma_0\langle \eta|\tilde{L}_0|\eta\rangle + \frac{1}{2} \Big\{ (\sigma_+-i\sigma_-)\langle\eta,h|\tilde{L}_1|\eta,h\rangle \\+(\sigma_++i\sigma_-)\langle \eta,h|\tilde{L}_{-1}|\eta,h\rangle \Big \}\Bigr],
\end{multline}
by making use of the identities
\begin{align}
\langle \eta|\tilde{L}_0|\eta\rangle&=h_q\frac{1+|\eta|^2}{1-|\eta|^2}+\text{cst},\\
\langle \eta|\tilde{L}_1|\eta\rangle &=\frac{2h_q\eta}{1-|\eta|^2},\\
\langle \eta|\tilde{L}_{-1}|\eta\rangle &=\frac{2h_q \bar{\eta}}{1-|\eta|^2}.
\end{align}
The classical Hamiltonian thus reads
\begin{equation}
\mathcal{H}  = \frac{2\pi q h_q}{L(1-|\eta|^2)}\Big[\sigma_0(1+|\eta|^2)+2\text{Re}\big\{(\sigma_+-i\sigma_-)\eta \big\}\Big].
\end{equation}
Using these ingredients, the classical equation of motion~\eqref{eq:classicalpoiusson} simplifies to a Riccati equation
\begin{equation}
\dot{\eta} = -\frac{2\pi i q}{L} \Big(\sigma_0 \eta + \frac{1}{2} \big[ (\sigma_++i\sigma_-)\eta^2+\sigma_+-i\sigma_-  \big] \Big).
\end{equation}
This equation can be solved analytically, leading to the Möbius transformation~\eqref{eq:transformationlaw}.

We now prove that the non-unitary evolution preserves coherent states, which is the statement of~\eqref{eq:coherentstatepresr}. In order to show this, we use the decomposition of coherent states as
\begin{equation}
    \ket{\eta,h} =  (1 - |\eta|^2)^{h_q} \sum_{n = 0}^{\infty} \eta^n \, \sqrt{\frac{\Gamma(2 h_q + n)}{\Gamma(n + 1)\Gamma(2 h_q)}} \, \ket{h, q, n} ,
    \label{eq:Coherent_state_explicit_form}
\end{equation}
where we defined the states
    $\ket{h, q, n} = \frac{1}{\sqrt{\mathcal N_{q,n}}} \tilde L_{-1}^n \ket{h}$, with normalization $\mathcal N_{q,n} = \frac{\Gamma(n + 1)\Gamma(2 h_q + n)}{\Gamma(2 h_q)}$. As a direct consequence, $\tilde{L}_0\ket{h,q,n} = (h_q+n)\ket{h,q,n}$. Therefore, it is clear that
    \begin{equation}
e^{- \tilde \lambda \tilde{L}_0}\ket{\eta,h} =\frac{e^{- \tilde \lambda h_q}(1-|\eta|^2)^{h_q}}{(1-e^{-2 \tilde \lambda}|\eta|^2)^{h_q}}|e^{- \tilde \lambda }\eta,h\rangle,
    \end{equation}
where we have defined the dimensionless quantity $\tilde \lambda \equiv 2 \pi \lambda / L$, and which immediately leads to~\eqref{eq:coherentstatepresr}.

\section{Displacement of primary operators}
\label{app:displacement_primaries}

In this appendix, we derive the differential equation~\eqref{eq:diff_eq_displacement} and show that its solution is~\eqref{eq:solution_differential_eq_displacement}. Using the Virasoro algebra, one has the following commutator with primary operators~\cite{DiFrancesco:639405}
\begin{multline}
    [\xi L_{- q} - \bar \xi L_q , \mathcal O(z)] = h \big[ \xi (1-q) z^{- q} - \bar \xi (1-q) z^{q} \big] \mathcal O(z) \\
    + \big( \xi z^{1 - q} - \bar \xi z^{1+q} \big) \partial_z \mathcal O(z) . 
\end{multline}
It is convenient to write the displacement operator as
\begin{equation}
    D(\xi) = e^{|\xi| (e^{i \theta} L_{-q} - e^{- i \theta} L_{q})}.
\end{equation}
Then, one has
\begin{multline}
    \partial_{|\xi|} \Big[ D^\dagger (\xi) \mathcal O(z) D (\xi) \Big] = \\
    h \big[e^{-i\theta} (1+q) z^{q} - e^{i\theta} (1-q) z^{- q} \big] \left(\frac{\partial f}{\partial z}\right)^{h}\mathcal{O}\big( f(\xi,z) \big) \\
    \big(e^{-i\theta} z^{1+q} - e^{i\theta} z^{1 - q} \big) \partial_z \left[ \left(\frac{\partial f}{\partial z}\right)^{h}\mathcal{O}\big( f(\xi,z) \big)  \right].
    \label{eq:LHS_displ_diff_eq}
\end{multline}
On the other hand, 
\begin{multline}
    \partial_{|\xi|} \left[ \left(\frac{\partial f}{\partial z}\right)^{h}\mathcal{O}\big( f(\xi,z) \big) \right] = \\
    h \left(\frac{\partial f}{\partial z}\right)^{h-1} \frac{\partial^2 f}{\partial z \partial |\xi|} \mathcal{O}\big( f(\xi,z) \big) + \\
    \left(\frac{\partial f}{\partial z}\right)^{h} \frac{\partial f}{\partial |\xi|} \partial_{f}\mathcal{O}\big( f(\xi,z) \big).
    \label{eq:RHS_displ_diff_eq}
\end{multline}
Equating~\eqref{eq:LHS_displ_diff_eq} and~\eqref{eq:RHS_displ_diff_eq} (respectively, the LHS and RHS of~\eqref{displacementeq}), we can organize the resulting equation into 
\begin{multline}
    h \partial_z \Big\{ \big[ e^{-i\theta} z^{1+q} - e^{i\theta} z^{1-q} \big] \partial_z f - \partial_{|\xi|} f\Big\} \mathcal{O}\big( f(\xi,z) \big) + \\
    \Big\{ \big[ e^{-i\theta} z^{1+q} - e^{i\theta} z^{1-q} \big] \partial_z f - \partial_{|\xi|} f\Big\} \partial_z f \, \partial_{f}\mathcal{O}\big( f(\xi,z) \big) = 0.
\end{multline}
This partial differential equation is solved by 
\begin{equation}
     \Big( e^{-i\theta} z^{1+q} - e^{i\theta} z^{1-q} \Big) \partial_z f - \partial_{|\xi|} f = 0 ,
\end{equation}
which is~\eqref{eq:diff_eq_displacement} advertised in the main text. In order to solve it, we can employ the change of coordinates
\begin{equation}
    w = -\frac{1}{q} \tanh^{-1} \big( e^{-i \theta} z^q \big),
\end{equation}
the equation above becomes
\begin{equation}
    \partial_w f - \partial_{|\xi|} f = 0 \ ,
\end{equation}
which is solved by
\begin{equation}
    f(\xi, w(z)) = F(w(z) + |\xi|) \ ,
\end{equation}
where $F$ is any differentiable function. However, for $\xi = 0$, the displacement operator is the identity, which demands $F(w(z)) = z$, fixing
\begin{equation}
    F(x) = \left( - e^{i \theta} \tanh(q x) \right)^{\frac{1}{q}}.
\end{equation}
All in all, we have
\begin{align}
    f(z, \xi) &= \left(e^{i \theta} \tanh \big(\tanh^{-1} \big( e^{-i \theta} z^q \big) - q |\xi| \big) \right)^{\frac{1}{q}} \\ 
    &= \left(\frac{ z^q - e^{i \theta} \tanh(q |\xi|)}{1 - e^{-i \theta} \tanh(q |\xi|) z^q}  \right)^{\frac{1}{q}},
\end{align}
which is~\eqref{eq:solution_differential_eq_displacement}.

\section{Fixed points for the periodic circuit}
\label{app:fixepoints}

In this appendix we give arguments that there is necessarily a stable fixed point in $\mathbb{D}$ for the periodic non-unitary drive discussed in Sec.~\ref{sec:periodicnonunitary21}, which involves as building blocks $e^{-\lambda H_0}$ and $e^{-i T H_1}$, with $H_1$ a general Hamiltonian of the form~\eqref{eq:defham}. Let us start with the simplest scenario, where the Floquet operator reads
\begin{equation}
U_{\text{F}} = e^{-\lambda H_0}e^{-iT H_1}.
\label{eq:firsttrial}
\end{equation}
In this case, the one-cycle coherent state is given by the SL$(2,\mathbb{C})$ transformation (here we set $L=2\pi$ for simplicity) 
\begin{equation}
 M_1 =  \begin{pmatrix}
 e^{-\lambda q} a & e^{-\lambda q}b \\
    e^{\lambda q}\bar b & e^{\lambda q} \bar a\\
\end{pmatrix},
\end{equation}
with $a$ and $b$ given by~\eqref{eq:mobiuscoef1} and~\eqref{eq:mobiuscoef2}, respectively. This transformation does not admit a fixed point if $|\text{Re}(\text{Tr}(M_1))|<2$ and $\text{Im}(\text{Tr}(M_1))=0$. It takes the form
\begin{equation}
\frac{1}{2}\text{Tr}(M_1) = \cos\left(\varphi\right)\cosh(\lambda q)-i\frac{\sigma_0}{\sqrt{C}}\sin\left(\varphi\right)\sinh(\lambda q),
\end{equation}
with $\varphi=\frac{q\pi \sqrt{C} T}{L}$.
Clearly as long as $\lambda\neq0$, $\text{Tr}(M_1)$ is real (i) if $\sigma_0=0$, which implies that $C<0$, such that $|\text{Tr}(M_1)|>2$, or (ii) if $\varphi=n\pi$, which also implies that $|\text{Tr}(M_1)|>2$. Therefore, we conclude that for a drive of the form~\eqref{eq:firsttrial}, there is necessarily a fixed point in $\mathbb{D}$. The physical consequence is that there always exists a steady state in the entanglement entropy and the energy for this non-unitary drive.

We can then consider more general one-cycle operators of the form
\begin{equation}
U_{\text{F}} = e^{-\lambda H_0}\prod_{j=1}^N e^{-i T_j H_j},
\end{equation}
for which the situation is the same, as we can reduce $\prod_{j=1}^N e^{-i T_j H_j} = e^{-i \tilde{H} \sum_{j=1}^N T_j}$, where  $\tilde{H}$ is obviously also of the form~\eqref{eq:defham}. This scenario thus simply reduces to the case studied previously.
It is nevertheless more complicated to argue that there will always be a stable fixed point for the most general Floquet operators, which ultimately reduce to the form 
\begin{equation}
U_{\text{F}}=\prod_{j=1}^Ne^{-\lambda_j H_0}e^{-i H_j T_j}.
\end{equation}
We have checked explicitly until $N=3$ that fixed points always arise as long as there is at least one $\lambda_j\neq0$.

\section{Fibonacci non-unitary circuits}
\label{app:quasiper}
In this appendix, we provide further detail on the transfer matrix approach to the Fibonacci quasiperiodic circuit defined by~\eqref{eq:fibonaccirecursion}. In particular, we argue that in the non-unitary setting, there cannot be any fractal parameter space, as opposed to the unitary setting studied in~\cite{PhysRevResearch.2.033461, PhysRevResearch.3.023044}. As discussed in Sec.~\ref{sec:fiboquasiperio}, the dynamics of coherent states can be reduced to a quasiperiodic product of matrices $M_0$ and $M_1$, following the recursion relation $M_n = M_{n-2}M_{n-1}$. For concreteness, we will pick
\begin{equation}
M_0 = \begin{pmatrix}
    e^{-\pi \lambda q/L}&0\\0& e^{\pi \lambda q/L}
\end{pmatrix}, \quad M_1 = \begin{pmatrix}
    a&b\\\bar b & \bar a
\end{pmatrix},
\end{equation}
with $a=\cosh(\pi qT/L)$, $b=i\sinh(\pi qT/L)$, which corresponds to $(\sigma_0,\sigma_+,\sigma_-)=(0,1,0)$ in~\eqref{eq:defham}.
Given that the matrix $M_n$ belongs in $\text{SL}(2,\mathbb{C})$, we can derive the following recursion formula for its trace using the Cayley–Hamilton theorem,
\begin{equation}
 x_{n+1} = 2x_nn_{n-1}-x_{n-2},
 \label{tracemapeq:}
\end{equation}
where we defined $x_n = \frac{1}{2}\Tr(M_n)$. By further defining  $y_n = \frac{1}{2}\Tr(M_{n+1})$ and $z_n = \frac{1}{2}\Tr(M_{n+2})$, we can rewrite this recursion relation as a map (called Fibonacci trace map)
\begin{equation}
\mathcal{T}: (x_n,y_n,z_n)\mapsto (y_n,z_n,2y_nz_n-x_n).
\label{eq:tracemap}
\end{equation}
In particular, note that $(x_0,y_0,z_0)\in\mathbb{R}^3$, and thus the trace map ~\eqref{tracemapeq:} defines a dynamical system over $\mathbb{R}^3$. The trace map~\eqref{eq:tracemap} is known to admit an invariant (called Fricke-Vogt invariant), 
\begin{equation}
 I = x_n^2+y_n^2+z_n^2-1-2x_ny_nz_n,
 \label{eq:frickeinvariant}
\end{equation}
which is independent of the Fibonacci step $n$.
In the unitary setting, this invariant was shown to always be nonnegative~\cite{PhysRevResearch.2.033461, PhysRevResearch.3.023044}, which leads to a fractal structure in parameter space, as it is known that for $I\geq0$ there is a Cantor set of orbits of the trace map~\eqref{eq:tracemap} which stay bounded at infinitely large times~\cite{Damanik_2008}. In the case at hand, it is straightforward to show that the invariant is strictly negative for a nonzero measurement strength,
\begin{equation}
 I = -\sinh^2(\pi qT/L)\sinh^2(\pi \lambda q/L) < 0.
\end{equation}
Then, the fractal set of bounded orbits does not exist, and any orbit starting from an initial point $(x_0,y_0,z_0)$ such that $||(x_0,y_0,z_0)||\geq \sqrt{3}$ is bound to diverge as $n\rightarrow\infty$ \footnote{We note that when $-1\leq I<0$, the invariant surface given by~\eqref{eq:frickeinvariant} admits an isolated compact component, which implies that all orbits with $||(x_0,y_0,z_0)||< \sqrt{3}$ are bounded~\cite{BAAKE_1993}. Nevertheless, in our case the norm of $(x_0,y_0,z_0)$ is always greater than $\sqrt{3}$ and thus all orbits diverge to infinity.}. Therefore, the fractal structure found in the unitary case~\cite{PhysRevResearch.2.033461, PhysRevResearch.3.023044} is absent in the non-unitary setting. Instead, the system always reaches a steady state at large times, independently of the initial state (although the value of the steady state depends on the parity of the Fibonacci step, as discussed in Sec.~\ref{sec:fiboquasiperio}).

\section{Lattice calculations}
\label{app:latcal}

The lattice calculations presented in the main text are based on free fermionic Hamiltonians of the form
\eqref{eq:latticemodel}. While similar results hold for open boundary conditions, we assume periodic boundary conditions. We consider as an initial state the ground state of the uniform chain with Hamiltonian $H_0$ at half-filling. Our aim is to calculate the time-evolved correlation matrix following the approach of~\cite{PhysRevX.13.021007}, from which we briefly summarize the results below. 

For simplicity, let us consider a two-step drive between the Hamiltonians $H_0$ and $H_1$, with Floquet operator
\begin{equation}
U_{\text{F}} = e^{-iT_1H_1}e^{-i(T_0-i\lambda)H_0}.
\end{equation}
We consider the initial state to be the ground state of $H_0$ at half-filling, written as
\begin{equation}
\ket{\Psi_0} = \prod_{i=1}^{L/2}\sum_{j=1}^L U_{ji}c_j^{\dagger}\ket{\text{vac}},
\label{eqgaussianinit}
\end{equation}
where we defined the unitary matrix $U$ that diagonalizes the single particle Hamiltonian, and $\ket{\text{vac}}$ is the fermionic vacuum.
Our goal is to compute the time-evolved correlation matrix after $n$ Floquet cycles,
\begin{equation}
C_{ij}(n) = \frac{\langle\Psi_0|(U^{\dagger}_{\text{F}})^n c_i^{\dagger}c_j (U_{\text{F}})^n|\Psi_0\rangle}{\langle\Psi_0|(U^{\dagger}_{\text{F}})^n (U_{\text{F}})^n|\Psi_0\rangle}.
\end{equation}
The latter can be computed explicitly given that the time evolution is quadratic and keeps the time-evolved state Gaussian,
\begin{equation}
\ket{\Psi(n)} = \prod_{i=1}^{L/2}\sum_{j=1}^L U_{ji}(n)c_j^{\dagger}\ket{\text{vac}},
\end{equation}
where we denote by $U(n)$ the $L\times\frac{L}{2}$ matrix that encodes the time-evolved Gaussian state after $n$ steps.  This matrix can be found directly by performing a $QR$ decomposition,
\begin{equation}
e^{-i T_1 H_1}e^{-i (T_0-i\lambda) H_0} U(n-1)=QR,
\end{equation}
where $Q$ is a symmetric $L\times L/2$ matrix satisfying $Q^{\dagger}Q=\mathbb{I}$, and $R$ is an upper triangular matrix.
The time evolution after $n$ steps is given by
\begin{equation}
U(n)=Q,
\end{equation}
and the initial condition $U(0)$ corresponds to the first $L/2$ columns of the matrix $U$ in~\eqref{eqgaussianinit}.
The correlation matrix after $n$ steps thus reads
\begin{equation}
C(n) = (U(n)U^{\dagger}(n))^T.
\end{equation}
From the correlation matrix $C(n)$ we can readily compute the von Neumann entanglement entropy using standard methods for free-fermionic lattice models~\cite{Peschel_2003}.

\bibliography{ref}

\end{document}